\documentclass[11pt]{article}
\usepackage[french,english]{babel}
\usepackage{amsmath}
\usepackage{amsfonts}
\usepackage{amssymb}
\usepackage{graphicx}        
\usepackage{times}
\begin{document}
%
\begin{titlepage}
June 15th  2010 (revised)  \hfill 
\newline\null
\vskip 6cm
{\baselineskip 17pt
\begin{center}
{\bf DISCRETE SYMMETRIES AND THE PROPAGATOR APPROACH

TO COUPLED FERMIONS IN QUANTUM FIELD THEORY .

 GENERALITIES .  THE CASE OF A SINGLE FERMION-ANTIFERMION PAIR .}
\end{center}
}
\vskip .2cm
\centerline{Q.~Duret
     \footnote[1]{LPTHE tour 13-14, 4\raise 3pt \hbox{\tiny \`eme} \'etage,
          UPMC Univ Paris 06, BP 126, 4 place Jussieu,
          F-75252 Paris Cedex 05 (France),\\
Unit\'e Mixte de Recherche UMR 7589 (CNRS / UPMC Univ Paris 06)}
    \footnote[2]{duret@lpthe.jussieu.fr}
           \& B.~Machet
    \footnotemark[1]
     \footnote[3]{machet@lpthe.jussieu.fr}
     }

\vskip 1.2cm
{\bf Abstract:}
 Starting from Wigner's symmetry representation theorem,
we give a general account of discrete symmetries (parity
$P$, charge conjugation $C$, time-reversal $T$),
focusing on fermions in Quantum Field Theory.
We provide the rules of transformation of Weyl spinors, both at the classical
level (grassmanian wave functions) and quantum level (operators).
Making use of Wightman's definition of invariance,
we outline ambiguities linked to the
notion of classical fermionic Lagrangian. We then present
the general constraints cast by these transformations 
and  their products on the  propagator of the simplest
among coupled fermionic system, the one made with one fermion and
its antifermion.  Last, we put in correspondence
the propagation of $C$ eigenstates (Majorana fermions) and
the criteria cast on their propagator by $C$ and $CP$ invariance.

\bigskip

{\it PACS: 11.10.Cd\quad 11.30.Cp\quad   11.30.Er}

{\it Keywords: spinors, Lorentz invariance,  discrete symmetries,
propagator}

\vfill
\end{titlepage}
%

\section{Introduction}
\label{section:introduction}

Fermions are usually treated, in most aspects of their phenomenology,
as classical, though anticommuting, objects.
Their Lagrangian is commonly endowed with a mass matrix though, for coupled systems
\footnote{Both quarks and leptons form  coupled systems through the
Higgs sector.},
this can only be a linear approximation in the vicinity of one among the
physical poles of their full (matricial) propagator \cite{MaNoVy}
 \cite{Novikov}. 
In this perspective, the study of neutral kaons \cite{MaNoVy},
and more specially of the role held, there, by discrete
symmetries $P$, $C$, $T$ and their products, has shown that subtle
differences occur between the ``classical'' treatment obtained from
a Lagrangian and a mass matrix, and the full quantum treatment dealing
with their propagator.
Using a classical approximation for fermions is {\em a
priori} still more subject to caution since, in particular, their
anticommutation is of quantum origin. This is why, after the work
\cite{MaNoVy}, we decided to perform a 
study of coupled fermionic systems in Quantum Field Theory,
dealing  especially with the propagator approach
\footnote{The propagator approach for coupled
systems was initiated in \cite{JacobSachs}, then applied in \cite{Sachs} 
to the case of neutral kaons. By defining the physical masses as the poles
of the full (matricial) propagator, it enabled to go beyond
the Wigner-Weisskopf approximation, to deal with non-hermitian Lagrangians
suitable for unstable particles,
and  to deduce general constraints cast  by discrete
symmetries. This method was then refined in \cite{MaNoVy}, still in the
case of neutral kaons.}.
Treating fermions in a rigorous way is all the more important as the very nature
of neutrinos, Dirac or Majorana, is still unknown, and that all theoretical
results, concerning specially flavor mixing,
have been mainly deduced from classical considerations.

The second and third parts of this work are dedicated to general statements
concerning, first, symmetry transformations in general, then
the discrete symmetries parity $P$, charge conjugation $C$, time-reversal
$T$, and their products. It does not pretend to be original, but tries to
make a coherent synthesis of results scattered in the literature.
Starting from Wigner's  representation theorem \cite{Wigner} and
Wightman's point of view for symmetry transformations \cite{Wightman},
 we give the general
rules of transformations of operators and of their hermitian conjugates
by any unitary or antiunitary transformation. We then specialize to
transforming Weyl spinors by $P$, $C$, $T$ and their products, first when
they are considered at the classical level (grassmanian wave functions),
then at the quantum level (anticommuting operators).

The fourth part deals with the concept of invariance of a given
theory.
By taking the simple example of fermionic mass
terms (Dirac and Majorana), we exhibit ambiguities and inconsistencies that
 arise in
the transformations of a classical Lagrangian by antiunitary
transformations. This motivates, like for neutral kaons \cite{MaNoVy},
the propagator approach, which is the only safe way of deducing unambiguously
the constraints cast by symmetry transformations on the Green functions of
physical (propagating) particles,  from which the
S-matrix can be in principle reconstructed \cite{Wightman}.

For the sake of simplicity, it is extensively
investigated only in the case of the simplest among coupled
fermionic systems, the one made with a single  fermion and
its antifermion; such a coupling, which concerns
neutral particles, is indeed allowed by Lorentz invariance.
This is the object of the fifth and last part of this work.
We derive in full
generality the constraints cast on the propagator
 by $P$, $C$, $T$, $PC$, $PCT$.  We show  that
the physical (propagating) fermions can only be Majorana ($C$ eigenstates) 
if their propagator satisfies the constraints cast by $C$ or
$CP$ invariance.

The  extension to several flavors,
with its expected deeper insight into the issue of quantum mixing
in connection with discrete symmetries, is currently under
investigation 
\footnote{Results concerning mixing at the quantum level 
have  been obtained, by  less general techniques, in
\cite{Beuthe}, \cite{DuMaVy} and \cite{DuMaVy2}.}.

\section{Generalities}
\label{section:generalities}

In this paper, we shall note equivalently
$\xi^\alpha \stackrel{\cal R}{\to} 
(\xi^\alpha)^{\cal R} \equiv {\cal R}\cdot\xi^\alpha $,
where $\xi^\alpha$ is a Weyl spinor (see Appendix \ref{subsection:spinors})
 and ${\cal R}\cdot\xi^\alpha$ its
transformed by $\cal R$; often the ``$\cdot$'' will be omitted such that
this transformed will also be noted ${\cal R}\xi^\alpha$.
The corresponding fermionic field operators will be put into square
brackets, for example $[\xi^\alpha], [\xi^\alpha]^{\cal R}$, the latter being the
transformed of the former  by the transformation $\cal R$.
Formally $ [\xi^\alpha]^{\cal R} = (\xi^\alpha)^{\cal R}$.

The transition amplitude between two fermionic states is noted $<\chi\ |\
\psi>$; this defines a scalar product and the corresponding norm
$<\psi\ |\ \psi>$ is real positive. The scalar product satisfies
\begin{equation}
<\psi\ |\ \chi>^\ast = <\chi\ |\ \psi> ;
\end{equation}
we consider furthermore  that representations of the
Poincar\'e group satisfy \cite{Wigner2}
\begin{equation}
<\psi\ |\ \chi>^\ast = <\psi^\ast\ |\ \chi^\ast>.
\end{equation}

\subsubsection{The symmetry representation theorem of Wigner \cite{Wigner}}
\label{subsub:wigner}

A symmetry transformation is defined as a transformation on the states (ray
representations) $\Psi \to \Psi'$ that preserve transition probabilities
\begin{equation}
|<\Psi'_1\ |\ \Psi'_2>|^2 = |<\Psi_1\ |\ \Psi_2>|^2.
\label{eq:symmetry}
\end{equation}

The so-called ``symmetry representation theorem'' states
\footnote{We refer  the reader to  \cite{Weinberg} for a careful
demonstration of this theorem.}:\newline
{\em Any symmetry transformation can be represented on the Hilbert space of
physical states by an operator that is either linear and unitary, or
antilinear and antiunitary.}

Since we have to deal with unitary  as well as antiunitary
operators, it is important to state their general properties and how they
operate on fermionic field operators.
A unitary operator $\cal U$ and an antiunitary operator $\cal A$ satisfy,
respectively
\begin{equation}
\forall \psi,\chi\quad <{\cal U}\psi\ |\ {\cal U}\chi>
= <\psi\ |\ \chi>,\quad
<{\cal A}\psi\ |\ {\cal A}\chi> = <\chi\ |\ \psi>
= <\psi\ |\ \chi>^\ast.
\label{eq:unianti1}
\end{equation}
Both preserve the probability transition $|<\psi\ |\ \chi>|^2 = |<{\cal U}\psi
\ |\
{\cal U}\chi>|^2 = |<{\cal A}\psi\ |\ {\cal A}\chi>|^2$.

\subsubsection{Antiunitarity and antilinearity}
\label{subsub:antianti}

An antilinear operator is an operator that complex conjugates any c-number
on its right
\begin{equation}
{\cal A}\ antilinear\ \Leftrightarrow\ 
{\cal A}\,(c\,|\ \psi>) = c^\ast {\cal A}\;|\ \psi>.
\label{eq:antilin}
\end{equation}
An antiunitary operator is also antilinear.
Let us indeed consider the antiunitary operator $\cal A$.

$<{\cal A} \psi \ |\ {\cal A}\ |\ \lambda \chi>
= <{\cal A}\psi\ |\ {\cal A}\lambda\chi>
= <\lambda\chi\ |\ \psi> = \lambda^\ast <\chi\ |\ \psi>
= \lambda^\ast <{\cal A}\psi\ |\ {\cal A}\ |\ \chi>$

shows that ${\cal A}$ is antilinear.

\subsubsection{Unitarity and linearity}
\label{subsub:unianti}

In the same way, one shows that a unitary operator is linear.

\subsubsection{Symmetry transformations: Wightman's point of view}
\label{subsub:symtrans}

Wightman \cite{Wightman} essentially deals with vacuum expectation
values of strings of field operators.
The transformed $\hat{\cal O}$ of an operator $\cal O$ is defined
through the transformation that changes the state $\phi$ into $\hat\phi$
\begin{equation}
<\hat\phi\ |\  {\cal O}\ |\ \hat\phi> = <\phi\ |\ \hat{\cal O}\ |\
\phi>
\label{eq:Wtrans}
\end{equation}
One has accordingly:\newline
* for a unitary transformation $\cal U$
\begin{equation}
\hat{\cal O}{=} {\cal U}^{-1}{\cal O}{\cal U},
\label{eq:opun}
\end{equation}
* for a antiunitary transformation $\cal A$
\footnote{The last equality in (\ref{eq:opanti}) comes from the
property, demonstrated by Weinberg \cite{Weinberg},
 that an antiunitary operator must also
satisfy the relation ${\cal A}{\cal A}^\dagger = 1 = {\cal A}^\dagger {\cal
A}$ (see Appendix
\ref{section:adjointanti}). So, in particular, one has
$({\cal A}^{-1})^\dagger {\cal A}^{-1}=1 \Rightarrow ({\cal A}^{-1})^\dagger =
{\cal A}$.\label{footnote:Weinberg}}
\footnote{
Because of (\ref{eq:opanti}), for ${\cal O} =
{\cal O}_1 {\cal O}_2\ldots {\cal O}_n$
\begin{eqnarray}
[{\cal O}_1{\cal O}_2\ldots {\cal O}_n]^\Theta &=&
\left( {\cal A}^{-1}{\cal O}_1{\cal O}_2\ldots {\cal O}_n{\cal A}\right)^\dagger
=\left( {\cal A}^{-1}{\cal O}_1{\cal A}{\cal A}^{-1}{\cal O}_2{\cal A}{\cal
A}^{-1}\ldots {{\cal A}{\cal A}^{-1}\cal O}_n{\cal A}\right)^\dagger\cr
&=&\left( {\cal A}^{-1}{\cal O}_n{\cal A}\right)^\dagger\ldots
\left( {\cal A}^{-1}{\cal O}_2{\cal A}\right)^\dagger
\left( {\cal A}^{-1}{\cal O}_1{\cal A}\right)^\dagger\cr
&=&[{\cal O}_n]^\Theta\ldots[{\cal O}_2]^\Theta[{\cal O}_1]^\Theta;
\label{eq:antiumit4}
\end{eqnarray}
antiunitarity implies that the order of operators has
to be swapped when calculating the transformed of a string of operators.
}
\begin{eqnarray}
\hat{\cal O}&=& ({\cal A}^{-1}{\cal O} {\cal A})^\dagger\cr
&=& {\cal A}^\dagger {\cal O}^\dagger ({\cal A}^{-1})^\dagger
= {\cal A}^\dagger {\cal O}^\dagger {\cal A}.
\label{eq:opanti}
\end{eqnarray}
This is the demonstration.\newline
* For $\cal U$ unitary (${\cal U}{\cal U}^\dagger = 1 = {\cal U}^\dagger{\cal
U}$):

$<{\cal U}\psi\ |\ {\cal O}\ |\ {\cal U}\chi> = <\psi\ |\ {\cal U}^\dagger{\cal O}{\cal
U}\ |\ \chi> = <\psi\ |\ {\cal U}^{-1}{\cal O}{\cal U}\ |\ \chi>,\ q.e.d.$

* For ${\cal A}$ antiunitary:\newline
- first, we demonstrate the important relation
\begin{equation}
\forall (\psi,\chi)\ <{\cal A}\,\psi\ |\ {\cal A}\, {\cal O}\, {\cal
A}^{-1}\ |\ {\cal A}\, \chi>
= <\chi\ |\ {\cal O}^\dagger\ |\ \psi>.
\label{eq:unianti2}
\end{equation}
Indeed:\newline
$<{\cal A} \psi\ |\  {\cal A} {\cal O}{\cal A}^{-1}\ |\ {\cal A} \chi>
= <{\cal A} \psi\ |\  {\cal A} {\cal O}\ |\ \chi> = <{\cal A} \psi\ |\ {\cal
A}({\cal O}\chi)>
\stackrel{(\ref{eq:unianti1})}{=}
<{\cal O}\chi\ |\ \psi> = <\chi\ |\  {\cal O}^\dagger\ |\ \psi>$;\newline
- one has then, in particular
\footnote{When the {\em in} and {\em out} states are different, one can
write accordingly
\begin{equation}
<{\cal A}\psi\ |\ {\cal O}\ |\ {\cal A}\chi> = <\chi\ |\ \hat {\cal O}\ |\ \psi>
= <\chi\ |\ ({\cal A}^{-1}{\cal O}{\cal A})^\dagger\ |\ \psi>
\label{eq:antiunit4}
\end{equation}
The {\em in} and {\em out} states have to be swapped in the expressions on
the r.h.s., ensuring
that all terms in (\ref{eq:antiunit4})  are linear in $\psi$ and
antilinear in $\chi$.}
\begin{equation}
<{\cal A}\, \psi\ |\  {\cal O}\ |\ {\cal A}\, \chi>
= <{\cal A} \, \psi \ |\ {\cal A} ({\cal A}^{-1} {\cal O}{\cal A}){\cal
A}^{-1}\ |\ {\cal A}\, \chi>
= <\chi\ |\ ({\cal A}^{-1} {\cal O}{\cal A})^\dagger\ |\ \psi>,
\label{eq:antiunit5}
\end{equation}
which yields the desired result for $\psi = \chi$
\footnote{One cannot use (\ref{eq:adjoint}) to transform $<\chi\ |\ ({\cal A}^{-1}
{\cal O}{\cal A})^\dagger\ |\ \psi>$ into $<\psi\ |\ {\cal A}^{-1} {\cal
O}{\cal A}\ |\ \chi$ because ${\cal A}^{-1} {\cal O}{\cal A}$ acts linearly
and should thus this considered as a unitary operator.}.

According to (\ref{eq:opanti}), an extra hermitian conjugation occurs in
the transformation of an operator by an anti-unitary transformation
\footnote{See \cite{Wightman}, eq.(1-30).}.

\subsubsection{General constraints}
\label{subsub:gencons}

$<\hat\phi\ |\ {\cal O}^\dagger \ |\ \hat\phi>
\stackrel{(\ref{eq:Wtrans})}{=}
<\phi\ |\ \widehat{{\cal O}^\dagger}\ |\ \phi>$ evaluates also as
$<\hat\phi\ |\ {\cal O}^\dagger \ |\ \hat\phi>
= <\hat\phi\ |\ {\cal O}\ |\ \hat\phi>^\ast
\stackrel{(\ref{eq:Wtrans})}{=} <\phi\ |\ \hat{\cal O}\ |\ \phi>^\ast\break
= <\phi\ |\ (\hat{\cal O})^\dagger\ |\ \phi>$, such that, comparing the two
expressions one gets
\begin{equation}
\widehat{{\cal O}^\dagger} =  (\hat{\cal O})^\dagger,
\label{eq:conjop}
\end{equation}
which is a constraint that must be satisfied by any operator $\cal O$
 transformed by unitary as well as antiunitary symmetry transformations.
Eq.~(\ref{eq:conjop}) can easily be checked explicitly.
$[\psi]$ being the field operator
associated with the grassmanian function $\psi$, one has:\newline
* for a unitary transformation $\cal U$: 
\begin{eqnarray}
\widehat{[\psi]^\dagger} & \stackrel{(\ref{eq:opun})}{=}&
 {\cal U}^{-1}\,[\psi]^\dagger\, {\cal U},\cr
\widehat{[\psi]^\dagger} &  \stackrel{(\ref{eq:conjop})}{=}&
 ([\hat\psi])^\dagger
\stackrel{(\ref{eq:opun})}{=} ({\cal U}^{-1}\,[\psi]\,{\cal U})^\dagger 
\stackrel{{\cal U}{\cal U}^\dagger =1={\cal U}^\dagger{\cal U}}{=}
{\cal U}^{-1}\,[\psi]^\dagger\, {\cal U};
\label{eq:tu}
\end{eqnarray}
* for a antiunitary transformation ${\cal A}$: 
\begin{eqnarray}
\widehat{[\psi]^\dagger}& \stackrel{(\ref{eq:opanti})}{=} &
({\cal A}^{-1}\,[\psi]^\dagger\, {\cal A})^\dagger
= {\cal A}^\dagger\, [\psi]\, {\cal A}, \cr
\widehat{[\psi]^\dagger}& \stackrel{(\ref{eq:conjop})}{=}&
({\cal A}^\dagger\, [\psi]^\dagger\, {\cal A})^\dagger
= {\cal A}^\dagger\, [\psi]\, {\cal A}.
\label{eq:ta}
\end{eqnarray}
Since $[\psi]$ and $[\psi]^\dagger$ are, respectively, associated with the
grassmanian functions $\psi$ and $\psi^\ast$, (\ref{eq:conjop}) 
also casts constraints on the transformation of grassmanian functions:
\begin{equation}
\widehat{\psi^\ast} = (\hat\psi)^\ast.
\label{eq:conjwave}
\end{equation}
%

\section{Discrete symmetries}
\label{section:discrete}

\subsection{Parity}
\label{subsection:parity}

\subsubsection{Parity transformation on grassmanian wave functions}

We adopt the convention  $P^2 = -1$ \cite{Landau}.
Then the transformation of spinors are
\begin{eqnarray}
\xi^\alpha(\vec x,t) \stackrel{P}{\to} i\eta_{\dot\alpha}(-\vec x,t)&,&
\eta_{\dot\alpha}(\vec x,t) \stackrel{P}{\to} i\xi^\alpha(-\vec x,t),\cr
\xi_\alpha(\vec x,t) \stackrel{P}{\to} -i\eta^{\dot\alpha}(-\vec x,t)&,&
\eta^{\dot\alpha}(\vec x,t) \stackrel{P}{\to} -i\xi_\alpha(-\vec x,t).
\label{eq:Pland1}
\end{eqnarray}
The parity transformed of the complex conjugates are defined \cite{Landau}
 as the complex conjugates of the parity transformed
\begin{equation}
P.(\xi^\alpha)^\ast = (P.\xi^\alpha)^\ast;
\end{equation}
this ensures in particular that the constraints (\ref{eq:conjop}) and
(\ref{eq:conjwave}) are satisfied.
It yields
\begin{eqnarray}
(\xi^\alpha)^\ast(\vec x,t) \stackrel{P}{\to}
-i(\eta_{\dot\alpha})^\ast(-\vec x,t)&,&
(\eta_{\dot\alpha})^\ast(\vec x,t) \stackrel{P}{\to}
-i(\xi^\alpha)^\ast(-\vec x,t),\cr
(\xi_\alpha)^\ast(\vec x,t) \stackrel{P}{\to}
i(\eta^{\dot\alpha})^\ast(-\vec x,t)&,&
(\eta^{\dot\alpha})^\ast(\vec x,t) \stackrel{P}{\to}
i(\xi_\alpha)^\ast(-\vec x,t).
\label{eq:Pland2}
\end{eqnarray}
For Dirac bi-spinors (see Appendix \ref{section:nota}), one gets
\begin{equation}
P.\psi_D = U_P \psi_D,\ U_P = i\gamma^0,
U_P^\dagger = -U_P=U_P^{-1}, U_P^2=-1, U_P^\dagger U_P=1.
\label{eq:UP}
\end{equation}
\subsubsection{Parity transformation on fermionic field operators}
\label{subsub:Pop}

Going to field operators, one uses (\ref{eq:opun}),
for unitary operators
\begin{equation}
[\xi^\alpha]^P = P^{-1} [\xi^\alpha] P
\end{equation}
to get
\begin{eqnarray}
P^{-1}\xi^\alpha(\vec x,t)P = i\eta_{\dot\alpha}(-\vec x,t)&,&
P^{-1}\eta_{\dot\alpha}(\vec x,t)P = i\xi^\alpha(-\vec x,t),\cr
P^{-1}\xi_\alpha(\vec x,t)P = -i\eta^{\dot\alpha}(-\vec x,t)&,&
P^{-1}\eta^{\dot\alpha}(\vec x,t)P = -i\xi_\alpha(-\vec x,t),\cr
P^{-1}(\xi^\alpha)^\dagger(\vec x,t)P =
-i(\eta_{\dot\alpha})^\dagger(-\vec x,t)&,&
P^{-1}(\eta_{\dot\alpha})^\dagger(\vec x,t)P =
-i(\xi^\alpha)^\dagger(-\vec x,t),\cr
P^{-1}(\xi_\alpha)^\dagger(\vec x,t)P =
i(\eta^{\dot\alpha})^\dagger(-\vec x,t)&,&
P^{-1}(\eta^{\dot\alpha})^\dagger(\vec x,t)P =
i(\xi_\alpha)^\dagger(-\vec x,t),\cr
&&
\label{eq:Plandop}
\end{eqnarray}
which satisfies the constraint (\ref{eq:conjop}).
The following constraint then arises
\begin{equation}
(P^{-1})^2 \xi^\alpha P^2 = -\xi^\alpha.
\label{eq:P2minus}
\end{equation}
Indeed: $(P^{-1})^2 \xi^\alpha P^2 =
P^{-1}(P^{-1} \xi^\alpha P)P \stackrel{(\ref{eq:Plandop})}{=}
P^{-1} i\eta_{\dot\alpha} P
\stackrel{linear}{=} i\, P^{-1}\eta_{\dot\alpha}P 
\stackrel{(\ref{eq:Plandop})}{=} -\xi^\alpha$.

Taking the hermitian conjugate of the first equation of the first line
 in (\ref{eq:Plandop}) and comparing it with the first equation of the
third line,
it is also immediate to check that $(P P^\dagger){\cal
O}(P P^\dagger)^{-1} = {\cal O},\ {\cal O} = \xi^\alpha \ldots,$
which is correct for $P$ unitary or antiunitary.

\subsection{Charge conjugation}
\label{subsec:C-lin}

$C$ is the operation which transforms a particle into its antiparticle, and
{\em vice versa}, without changing its spin and momentum (see for example
\cite{BrancoLavouraSilva} p.17);
it satisfies $C^2 = 1$ \cite{Landau}

\subsubsection{Charge conjugation of grassmanian wave functions}

A Dirac fermion and its charge conjugate transform alike \cite{Landau}
and satisfy the same equation; the charge conjugate satisfies
\begin{equation}
C\cdot\psi_D = V_C \overline{\psi_D}^T,
\label{eq:Cdef}
\end{equation}
where $V_C$ is a  unitary operator
\begin{equation}
V_C = \gamma^2\gamma^0,\quad (V_C)^\dagger V_C =1= (V_C)^2;
\end{equation}
 equivalently
\begin{equation}
C\cdot\psi_D  = U_C \psi_D^\ast,\quad U_C =V_C\gamma^0= \gamma^2,
\ U_C^\dagger U_C =1=-(U_C)^2.
\label{eq:UC}
\end{equation}
In terms of Weyl fermions (see Appendix \ref{section:nota}), one has
\begin{equation}
\psi_D \equiv\left(\begin{array}{c} \xi^\alpha \cr \eta_{\dot\beta}\end{array}\right)
\stackrel {C}{\to}
-i\left(\begin{array}{c} \eta^{\dot\alpha\ast} \cr \xi_{\beta}^\ast
\end{array}\right)
=-i\left(\begin{array}{c} g^{\dot\alpha\dot\beta}\eta_{\dot\beta}^\ast
\cr g_{\alpha\beta}\xi^{\beta\ast}
\end{array}\right)
=\left(\begin{array}{r} -\sigma^2_{\dot\alpha\dot\beta}\eta_{\dot\beta}^\ast
\cr \sigma^2_{\alpha\beta}\xi^{\beta\ast}
\end{array}\right)
=\gamma^2 \left(\begin{array}{c}
\xi^\alpha \cr \eta_{\dot\beta}\end{array}\right)^\ast
=\gamma^2 \psi_D^\ast,
\label{eq:Cdirac}
\end{equation}
and, so
\begin{eqnarray}
\xi^\alpha \stackrel{C}{\to} -i \eta^{\dot\alpha\ast}&,&
\eta_{\dot\alpha}\stackrel{C}{\to} -i \xi_\alpha^\ast,\cr
\xi_\alpha \stackrel{C}{\to} -i \eta_{\dot\alpha}^\ast&,&
\eta^{\dot\alpha}\stackrel{C}{\to} -i \xi^{\alpha^\ast}.
\label{eq:Cland1}
\end{eqnarray}
The transformation of complex conjugates fields results from the
 constraint (\ref{eq:conjwave}), which imposes
\begin{eqnarray}
(\xi^\alpha)^\ast \stackrel{C}{\to}
i \eta^{\dot\alpha}&,&
(\eta_{\dot\alpha})^\ast\stackrel{C}{\to}
i \xi_\alpha,\cr
(\xi_\alpha)^\ast \stackrel{C}{\to}
i \eta_{\dot\alpha}&,&
(\eta^{\dot\alpha})^\ast\stackrel{C}{\to}
i \xi^{\alpha}.
\label{eq:Cland1c}
\end{eqnarray}
One can now show that (recall that $U_C^2 = -1$ from (\ref{eq:UC}))
\begin{equation}
C\ unitary\ and\ linear, \ C^2=1.
\label{eq:clin}
\end{equation}
If (\ref{eq:conjwave}) holds, the property $C^2=1$ can
only be realized if one considers that
$C$ is a linear operator. Indeed, then, using (\ref{eq:Cland1}) and
(\ref{eq:Cland1c}), one has
$C\cdot C\cdot \xi^\alpha \stackrel{(\ref{eq:Cland1})}{=} C\cdot(-i(\eta^{\dot\alpha})^\ast)
\stackrel{linear}{=} (-i) C\cdot(\eta^{\dot\alpha})^\ast
\stackrel{(\ref{eq:Cland1c})}{=}\xi^\alpha$, which entails, as needed,
 $C^2=1$.

The only way to keep $C^2=1$ while having $C$ antilinear, as (\ref{eq:UC})
seems to suggest,  would be to break the relation  (\ref{eq:conjwave}),
 in which case, the signs of (\ref{eq:Cland1c}) get swapped.
Suppose indeed that we consider that $C$ is
antilinear (thus also antiunitary), and suppose that we also want 
to preserve the relation (\ref{eq:conjwave}); then,
(\ref{eq:Cland1c}) stays true together with (\ref{eq:Cland1}), 
and, by operating a second time
with $C$ on the l.h.s. of (\ref{eq:Cland1}) or (\ref{eq:Cland1c}),
one finds that it can only satisfy $C^2=-1$ instead of $C^2=1$. Among
consequences, one finds that
the commutation and anticommutation relations with other symmetry
transformations $P$ and $T$ are  changed
\footnote{With our conventions, we have  $CP=PC, (PC)^2=-1$, and $(PCT)^2 =1$.},
 which  swaps in particular the sign of $(PCT)^2$;
also, since $T$ is antilinear and $P$ is linear, this would make $PCT$
linear, thus unitary.
So, if we want $C$ to be antilinear, we have to abandon
(\ref{eq:conjwave}); considering that, at the same time, the equivalent
relation (\ref{eq:conjop}) for operators in not true either causes serious
problems with Wightman's definition (\ref{eq:Wtrans})
of the transformed of an operator (see subsection \ref{subsub:gencons})
which has to be either unitary or antiunitary according to the  Wigner's
symmetry representation theorem (see subsection \ref{subsub:wigner}).
Refusing to go along this path, 
we have to keep (\ref{eq:conjop}) while giving up
(\ref{eq:conjwave}), that is we must abandon the natural correspondence
$\psi \leftrightarrow [\psi], \psi^\ast \leftrightarrow [\psi]^\dagger$
between fields and operators. This looks extremely unnatural and a price
too heavy to pay; this is why 
we consider that the relations (\ref{eq:conjwave}) and $C^2=1$ are 
only compatible with unitarity and linearity for $C$.

The question now arises whether this causes any problem or leads to
contradictions, thinking in
particular of  (\ref{eq:Cdef}) and (\ref{eq:UC}); if one indeed considers these
two equations as the basic ones defining charge conjugation, one is  led to
$C\cdot(\lambda \psi_D) =\lambda^\ast C\cdot(\psi_D)$ and
that, accordingly, $C$ acts antilinearly on wave functions.
Our argumentation rests on the fact that  (\ref{eq:Cdef}) and (\ref{eq:UC})
should not be considered as so. Indeed, the two conditions
defining the action of $C$ are \cite{Landau}; -- that a fermion and its
charge conjugate should transform alike by Lorentz; -- that they should
satisfy the same equation.  Since the Dirac equation is
linear, both $\lambda C\cdot\psi_D$ and $\lambda^\ast C\cdot\psi_D$ satisfy the
same Dirac equation as $C\cdot\psi_D$, and thus, the same equation as $\psi_D$.
Likewise, both $\lambda C\cdot\psi_D$ and $\lambda^\ast C\cdot\psi_D$ transform by
Lorentz as $C\cdot\psi_D$, and thus, as $\psi_D$. So, the two fundamental
requirements concerning the charge conjugate of a Dirac fermion bring no
constraint on the linearity or antilinearity of $C$, and this last property
must be fixed by other criteria. The ones in favor of a linear action of $C$
have been enumerated above:
-- to preserve the relation $C^2=1$;
-- to preserve  Wightman's definition of a symmetry
transformation and to stick to Wigner's symmetry representation theorem;
-- to preserve both relations (\ref{eq:conjwave}) and (\ref{eq:conjop});
-- to preserve the natural correspondence between wave functions
and field operators.
Our final proposition is accordingly that:
despite $C$ complex conjugates a Dirac spinor, it has to be considered as a
linear and unitary operator  (in particular the relation
 $C\cdot\lambda\psi = \lambda\, C\cdot\psi$ has to be imposed),
and this does not depend on whether it acts on a wave function or
on a field operator.

We also refer the reader to appendix \ref{section:becareful}, where
a careful analysis is done
of the pitfalls that accompany the use of $\gamma$ matrices in the
expression of the discrete transformations $P$, $C$ and $T$.

\subsubsection{Charge conjugation of fermionic field operators}
\label{subsub:Cop}

According to the choice of linearity and unitarity for $C$,
the transition from (\ref{eq:Cland1}) and (\ref{eq:Cland1c}) for
grassmanian wave
functions to the transformations for field operators is done according
to (\ref{eq:opun}) for  unitary operators, through the correspondence
${\cal U}\psi \leftrightarrow {\cal U}^{-1} [\psi]\, {\cal U}$. One gets
\begin{eqnarray}
C^{-1} \xi^\alpha C =  -i (\eta^{\dot\alpha})^\dagger &,&
C^{-1} \eta_{\dot\alpha} C= -i (\xi_\alpha)^\dagger,\cr
C^{-1} \xi_\alpha C = -i (\eta_{\dot\alpha})^\dagger &,&
C^{-1} \eta^{\dot\alpha} C =-i (\xi^{\alpha})^\dagger,\cr
C^{-1} (\xi^\alpha)^\dagger C =i (\eta^{\dot\alpha})&,&
C^{-1} (\eta_{\dot\alpha})^\dagger C =i (\xi_\alpha),\cr
C^{-1} (\xi_\alpha)^\dagger C =i (\eta_{\dot\alpha})&,&
C^{-1} ((\eta^{\dot\alpha})^\dagger C =i (\xi^{\alpha}).
\label{eq:psiopC}
\end{eqnarray}
Hermitian conjugating the first equation of the first line of
(\ref{eq:psiopC}) immediately shows its compatibility with
 the first equation of the third line:
$C^\dagger (\xi^\alpha)^\dagger (C^{-1})^\dagger = i
\eta^{\dot\alpha} =C^{-1}(\xi^\alpha)^\dagger C \Rightarrow
(\xi^\alpha)^\dagger = C C^\dagger (\xi^\alpha)^\dagger
(C^{-1})^\dagger C^{-1}$, which entails $C C^\dagger = \pm 1$ which
is correct for $C$ unitary (or antiunitary). We would find an
inconsistency if the sign of the last four equations was swapped.

Since $C$ is linear, one immediately gets
\begin{equation}
(C^{-1})^2\, {\cal O}\, C^2
= C^{-1}(C^{-1}\, {\cal O}\, C)C
= {\cal O}, {\cal O} = \xi^\alpha
\ldots
\end{equation}

\subsection{$\boldsymbol{PC}$ transformation}
\label{subsection:CPtrans}

\subsubsection{$\boldsymbol{PC}$ transformation on grassmanian wave
functions}

Combining (\ref{eq:Pland1}), (\ref{eq:Cland1}) and (\ref{eq:Cland1c}),
and using, when needed, the linearity of $C$, one gets
\begin{eqnarray}
\xi^\alpha(\vec x,t) \stackrel{PC}{\to} \xi_\alpha^\ast(-\vec x,t)&,&
\eta_{\dot\alpha}(\vec x,t) \stackrel{PC}{\to} \eta^{\dot\alpha\ast}(-\vec
x,t),\cr
\xi_\alpha(\vec x,t) \stackrel{PC}{\to} -\xi^{\alpha^\ast}(-\vec x,t)
&,&
\eta^{\dot\alpha}(-\vec x,t) \stackrel{PC}{\to}
 -\eta_{\dot\alpha}^\ast(-\vec x,t),
\label{eq:CPlag1}
\end{eqnarray}
and

\vbox{
\begin{eqnarray}
(\xi^\alpha)^\ast(\vec x,t) \stackrel{PC}{\to} \xi_\alpha(-\vec x,t) &,&
(\eta_{\dot\alpha})^\ast(\vec x,t) \stackrel{PC}{\to} \eta^{\dot\alpha}(-\vec
x,t),\cr
(\xi_\alpha)^\ast(\vec x,t) \stackrel{PC}{\to} -\xi^\alpha(-\vec x,t)
&,&
(\eta^{\dot\alpha})^\ast(\vec x,t) \stackrel{PC}{\to} -\eta_{\dot\alpha}(-\vec x,t).
\label{eq:CPlag2}
\end{eqnarray}
}

One easily checks that $(PC)^2 =-1$.

Like for charge conjugation, one has
\begin{equation}
PC\cdot(\xi^\alpha)^\ast = \big(PC\cdot\xi^\alpha\big)^\ast.
\end{equation}
For a Dirac fermion,  one has
\begin{equation}
\left(\begin{array}{c} \xi^\alpha \cr \eta_{\dot\beta}\end{array}\right)
\stackrel{PC}{\to}
\left(\begin{array}{c} \xi_\alpha^\ast \cr \eta^{\dot\beta\ast}\end{array}
\right)
=\left(\begin{array}{c} g_{\alpha\beta}\xi^{\beta\ast} \cr
g^{\dot\beta\dot\gamma}\eta_{\dot\gamma}^\ast\end{array}
\right)
=\left(\begin{array}{c} (i\sigma^2)_{\alpha\beta}\xi^{\beta\ast} \cr
(-i\sigma^2)_{\dot\beta\dot\gamma}\eta_{\dot\gamma}^\ast\end{array}
\right)
=
i\left(\begin{array}{c} (\eta_{\dot\alpha})^c \cr (\xi^{\beta})^c\end{array}
\right) = i\gamma^0\gamma^2 \left(\begin{array}{c}\xi^\alpha \cr
\eta_{\dot\beta}\end{array}\right)^\ast,
\end{equation}
equivalently
\begin{equation}
PC\cdot \psi_D = V_{PC} \overline{\psi}^T = U_P V_C \overline{\psi}^T =
 U_{PC} \psi^\ast = U_P U_C \psi^\ast.
\end{equation}

As we will see in subsection \ref{subsection:majorana}, Majorana fermions
 have $PC$-{\em parity} $\pm i$.

\subsubsection{$\boldsymbol{PC}$ transformation on fermionic field
operators}

Since we have defined $PC$ as a linear (and unitary) operator, the
transitions from grassmanian wave functions to field operators goes
through  (\ref{eq:opun}). This yields
\begin{eqnarray}
(PC)^{-1} \xi^\alpha (PC) =  \xi_\alpha^\dagger &,&
(PC)^{-1} \eta_{\dot\alpha} (PC)=  (\eta^{\dot\alpha})^\dagger,\cr
(PC)^{-1} \xi_\alpha (PC) = - (\xi^\alpha)^\dagger &,&
(PC)^{-1} \eta^{\dot\alpha} (PC) =- (\eta_{\dot\alpha})^\dagger,\cr
(PC)^{-1} (\xi^\alpha)^\dagger (PC) = \xi_\alpha&,&
(PC)^{-1} (\eta_{\dot\alpha})^\dagger (PC) = \eta^{\dot\alpha},\cr
(PC)^{-1} (\xi_\alpha)^\dagger (PC) = -\xi^\alpha&,&
(PC)^{-1} ((\eta^{\dot\alpha})^\dagger (PC) = - \eta_{\dot\alpha}.
\label{eq:CPop}
\end{eqnarray}

\subsection{Time-reversal}
\label{subsection:timerev}

\subsubsection{Time-reversal of grassmanian wave functions}

The time reversed $<\chi(t')\ |\ \psi(t)>^T$
of a transition matrix element $<\chi(t')\ |\ \psi(t)>,
t<t'$ is defined by
 $<\chi(t)\ |\ \psi(t')>^\ast = <\psi(t')\ |\
\chi(t)>, t>t'$; the complex conjugation is made necessary
by $t<t'$ and the fact that  {\em in} states must occur at a time smaller
than {\em out} states; the arrow of time is not modified when one
defines the time-reversed of a transition matrix element.

The operator $T$ is accordingly antiunitary, hence antilinear:
\begin{equation}
<TA\ |\ TB> = <B\ |\ A> \Rightarrow T\ antiunitary,
\end{equation}
In Quantum Mechanics, time-reversal must  change grassmanian functions into
their complex conjugate (see for example the argumentation concerning
Schr{\oe}dinger's equation in \cite{BrancoLavouraSilva}).
According to \cite{Landau}, the {\em grassmanian functions} transform by
time inversion according to
\begin{eqnarray}
&&\psi_D(\vec x,t) \stackrel{T}{\to} T\cdot\psi_D(\vec x,t)
= V_T \overline{\psi_D(\vec x,-t)}^T;\cr
&&V_T = i\gamma^3\gamma^1\gamma^0,\quad V_T^\dagger V_T =1 = V_T^2,
\quad V_T^\dagger = V_T = V_T^{-1},
\end{eqnarray}
which introduces $T$ as  antilinear when it acts on grassmanian functions.
So doing, $T.\psi_D$ and $\psi_D$ satisfy time reversed equations.
One also defines
\begin{equation}
U_T = V_T \gamma^0 = i\gamma^3\gamma^1=-U_T^\ast, U_T^\dagger =
 U_T =U_T^{-1}, U_T^\dagger U_T = U_T^2= 1.
\label{eq:UT}
\end{equation}
\begin{equation}
T\cdot\psi_D = U_T \psi_D^\ast = i\gamma^3\gamma^1 \psi_D^\ast.
\label{eq:Ut1}
\end{equation}
This yields for Weyl fermions
\begin{eqnarray}
\xi^\alpha(\vec x,t) \stackrel{T}{\to}
 -i \xi_\alpha^\ast(\vec x,-t)&,&
\xi_\alpha(\vec x,t) \stackrel{T}{\to}
i\xi^{\alpha\ast}(\vec x,-t),\cr
\eta_{\dot\alpha}(\vec x,t) \stackrel{T}{\to}
i\eta^{\dot\alpha\ast}(\vec x,-t)&,&
\eta^{\dot\alpha}(\vec x,t) \stackrel{T}{\to}
-i \eta_{\dot\alpha}^\ast(\vec x,-t).
\label{eq:Tland1}
\end{eqnarray}
The constraint (\ref{eq:conjwave}) then entails
\begin{eqnarray}
 (\xi^\alpha)^\ast(\vec x,t) \stackrel{T}{\to} i\xi_\alpha(\vec x,-t) &,&
 (\xi_\alpha)^\ast(\vec x,t) \stackrel{T}{\to} -i\xi^\alpha(\vec x,-t) ,\cr
(\eta_{\dot\alpha})^\ast(\vec x,t) \stackrel{T}{\to}
-i\eta^{\dot\alpha}(\vec x,-t)&,&
(\eta^{\dot\alpha})^\ast(\vec x,t) \stackrel{T}{\to}
i\eta_{\dot\alpha}(\vec x,-t).
\label{eq:Tland3}
\end{eqnarray}
One has
\begin{equation}
T^2=1,\  CT = -TC,\ PT=TP
\end{equation}

\subsubsection{Time-reversal of fermionic field operators}
\label{subsub:timeop}

The transition to field operators is done according to (\ref{eq:opanti})
for antiunitary transformations, through the correspondence
 $({\cal A}\psi)^\dagger \leftrightarrow {\cal A}^{-1} [\psi]{\cal A}$,
which involves an extra hermitian
conjugation with respect to the transformations of grassmanian functions
(\cite{Wightman}, eq.(1-30)):
\begin{eqnarray}
T^{-1}\xi^\alpha(\vec x,t) T=
 i \xi_\alpha(\vec x,-t)&,&
T^{-1}\eta_{\dot\alpha}(\vec x,t)T=
-i\eta^{\dot\alpha}(\vec x,-t),\cr
T^{-1}\xi_\alpha(\vec x,t)T=
-i\xi^\alpha(\vec x,-t) &,&
T^{-1}\eta^{\dot\alpha}(\vec x,t)T=
i \eta_{\dot\alpha}(\vec x,-t),\cr
T^{-1} (\xi^\alpha)^\dagger(\vec x,t) T=
-i(\xi_\alpha)^\dagger(\vec x,-t) &,&
T^{-1} (\xi_\alpha)^\dagger(\vec x,t) T=
i(\xi^\alpha)^\dagger(\vec x,-t) ,\cr
T^{-1}(\eta_{\dot\alpha})^\dagger(\vec x,t) T=
i(\eta^{\dot\alpha})^\dagger(\vec x,-t)&,&
T^{-1}(\eta^{\dot\alpha})^\dagger(\vec x,t) T=
-i(\eta_{\dot\alpha})^\dagger(\vec x,-t).
\label{eq:Top3}
\end{eqnarray}
Since $T$ is antilinear, one finds immediately that, though
$T^2 = 1$, one must have
\begin{equation}
(T^{-1})^2\, {\cal O}\, T^2
=T^{-1}(T^{-1}\, {\cal O}\, T)T
=  -{\cal O}, {\cal O} =
\xi^\alpha \ldots
\end{equation}

\subsection{$\boldsymbol{PCT}$ transformation}
\label{subsection:TCPtransW}

\subsubsection{$\boldsymbol{PCT}$ operation on grassmanian wave functions}

Combining the previous results, using the linearity of $P$ and $C$,
 one gets {\em for the grassmanian functions}
\footnote{Examples:

$PCT\cdot \xi^\alpha = PC\cdot(T\cdot\xi^\alpha) =
PC\cdot(-i \xi_\alpha^\ast) = P\cdot (-i)C\cdot\xi_\alpha^\ast = (-i)P\cdot C\cdot\xi_\alpha^\ast
=(-i)P\cdot i\eta_{\dot\alpha} = P\cdot\eta_{\dot\alpha} = i\xi^\alpha$;

$PCT\cdot (\xi^\alpha)^\ast =PC\cdot(T\cdot(\xi^\alpha)^\ast))
=PC\cdot (i\xi_\alpha) = P\cdot i C\cdot\xi_\alpha =
iP\cdot (-i)(\eta_{\dot\alpha})^\ast = P\cdot (\eta_{\dot\alpha})^\ast =
-i(\xi^\alpha)^\ast$.
}

\begin{eqnarray}
\xi^\alpha(x) \stackrel{PCT}{\to} i\xi^\alpha(-x)&,&
\eta_{\dot\alpha}(x) \stackrel{PCT}{\to} -i\eta_{\dot\alpha}(-x),\cr
\xi_\alpha(x) \stackrel{PCT}{\to} i\xi_\alpha(-x)&,&
\eta^{\dot\alpha}(x) \stackrel{PCT}{\to} -i\eta^{\dot\alpha}(-x),\cr
&& \hskip -2cm\psi_D(x) \stackrel{PCT}{\to} i\gamma^5 \psi_D(-x),
\label{eq:TCPland1b}
\end{eqnarray}
where the overall sign depends on the order in which the operators act;
here they are supposed to act in the order: first $T$,
then $C$ and last $P$.
When acting on bi-spinors, one has $CT = -TC$ and $PT=TP$
\footnote{We disagree with  \cite{Landau}
 who states that $T$ and $P$ anticommute.}.
So, using also $CP=PC$, one gets
$(PCT)(PCT) = (PCT)(P(-)TC)=  (PCT)(-TPC)$. $T^2 =1$, $C^2=1$, $P^2=-1$
(our choice) and $PC=CP$ entail
\begin{equation}
(PCT)^2 =1.
\end{equation}
Note that, both $C$ and $T$  introducing complex
conjugation, the latter finally disappears and
${PCT}$ introduces  no complex conjugation for the grassmanian functions.
This is why one has
\begin{equation}
PCT\cdot \psi_D(x) = U_\Theta \psi_D(-x),
\end{equation}
\begin{equation}
U_\Theta = U_P U_C U_T=-\gamma^0\gamma^1\gamma^2\gamma^3 = i\gamma^5,\quad
U_\Theta U_\Theta^\dagger =1 = -U_\Theta^2, \quad
U_\Theta^\dagger=-U_\Theta.
\label{eq:Utheta}
\end{equation}
For the complex conjugate fields, the constraint (\ref{eq:conjwave}) gives
\begin{eqnarray}
(\xi^\alpha)^\ast(x) \stackrel{PCT}{\to} -i(\xi^\alpha)^\ast(-x)&,&
(\eta_{\dot\alpha})^\ast(x) \stackrel{PCT}{\to} i(\eta_{\dot\alpha})^\ast(-x),\cr
(\xi_\alpha)^\ast(x) \stackrel{PCT}{\to} -i(\xi_\alpha)^\ast(-x)&,&
(\eta^{\dot\alpha})^\ast(x) \stackrel{PCT}{\to} i(\eta^{\dot\alpha})^\ast(-x),\cr
&& \hskip -2cm\psi_D^\ast \stackrel{PCT}{\to} -i\gamma^5 \psi_D^\ast,
\label{eq:TCPland2b}
\end{eqnarray}
such that (this only occurs for $P$ and $PCT$)
\begin{equation}
PCT\cdot(\xi^\alpha)^\ast =  (PCT\cdot\xi^\alpha)^\ast \Leftrightarrow
U_\Theta (\xi^\alpha)^\ast \equiv ((\xi^\alpha)^\ast)^\Theta =
(U_\Theta \xi^\alpha)^\ast \equiv ((\xi^\alpha)^\Theta)^\ast.
\end{equation}
Since $P$ and $C$ are unitary and $T$
antiunitary,  $PCT$  is antiunitary, thus {\em antilinear}.
So, despite no complex conjugation is involved
$\Theta\cdot\lambda\xi^\alpha = \lambda^\ast \Theta\cdot\xi^\alpha$
\footnote{This is to be put in correspondence with $C$, which
is {\em linear} despite complex conjugation is involved.}.

\subsubsection{$\boldsymbol{PCT}$ operation on fermionic field operators}
\label{subsub:PCTop}

Since $\Theta$ is antiunitary, one has, according to (\ref{eq:opanti})
\begin{eqnarray}
\Theta^{-1} \xi^\alpha(x) \Theta = -i(\xi^\alpha)^\dagger(-x)  &,&
\Theta^{-1} \xi_\alpha(x) \Theta =  -i(\xi_\alpha)^\dagger(-x),\cr
\Theta^{-1} \eta_{\dot\alpha}(x) \Theta =
i(\eta_{\dot\alpha})^\dagger(-x) &,&
\Theta^{-1} \eta^{\dot\alpha}(x) \Theta =
i(\eta^{\dot\alpha})^\dagger(-x),\cr
\Theta^{-1} (\xi^\alpha)^\dagger(x) \Theta = i\xi^\alpha(-x) &,&
\Theta^{-1} (\xi_\alpha)^\dagger(x) \Theta =i\xi_\alpha(-x) ,\cr
\Theta^{-1} (\eta_{\dot\alpha})^\dagger(x) \Theta =
-i\eta_{\dot\alpha}(-x) &,&
\Theta^{-1} (\eta^{\dot\alpha})^\dagger(x) \Theta =
-i\eta^{\dot\alpha}(-x).
\label{eq:TCPop}
\end{eqnarray}

and, using the antilinearity of $\Theta$, one gets
\begin{equation}
(\Theta^{-1})^2 \, {\cal O}\, \Theta^2
= \Theta (\Theta^{-1} \, {\cal O}\, \Theta)\Theta
= -{\cal O}, {\cal O} = \xi^\alpha \ldots
\end{equation}

\subsection{Majorana fermions}
\label{subsection:majorana}

A Majorana fermion is a bi-spinor which is a $C$ eigenstate
(it is  a special kind of Dirac fermion with half as many
degrees of freedom); since $C^2=1$, the only two possible eigenvalues are
$C=+1$ and $C=-1$; thus, a Majorana fermions must satisfy
(see (\ref{eq:Cdirac})) one of the two possible  Majorana conditions:

$\ast$\  $-i \eta^{\dot\alpha\ast} = \pm \xi^\alpha
\Leftrightarrow \eta^{\dot\alpha} = \pm(-i)\xi^{\alpha\ast} \Leftrightarrow
\eta_{\dot\beta} = \pm (-i) \xi_\beta^\ast$;

$\ast$\ $-i\xi_\beta^\ast = \pm \eta_{\dot\beta}$, which is 
the same condition as above;

so,
\begin{equation}
\psi_{M}^\pm = \left(\begin{array}{c} \xi^\alpha \cr \pm(-i) \xi_\beta^\ast
\end{array}\right)
= \left(\begin{array}{c} \xi^\alpha \cr \pm(-i)
g_{\alpha\beta}\xi^{\beta\ast}\end{array}\right)
= \left(\begin{array}{c} \xi^\alpha \cr \pm
\sigma^2_{\alpha\beta}\xi^{\beta\ast}\end{array}\right);
\label{eq:psiMaj}
\end{equation}
the $+$ sign in the lower spinor corresponds to $C=+1$ and the $-$ sign to
$C=-1$
\footnote{{\em Remark}: Arguing that $(-i)(\xi_\beta)^\ast)$ transforms like a
right fermion, we can call $\omega_{\dot\beta} =(-i)(\xi_\beta)^\ast)$, and
the Majorana fermion $\psi^+_M$  rewrites
$\psi^+_M= \left(\begin{array}{c}\xi^\alpha \cr
\omega_{\dot\beta}\end{array}\right)$. If we then calculate its charge
conjugate according to the standard rules (\ref{eq:Cland1}), one gets
$\psi^+_M \stackrel{C}{\to} \left(\begin{array}{c}
-i (\omega^{\dot\alpha})^\dagger \cr -i(\xi_\alpha)^\ast\end{array}\right)
\equiv \left(\begin{array}{c} \xi^\alpha \cr
-i(\xi_\alpha)^\ast\end{array}\right)$, which shows that it is indeed a $C=+1$
eigenstate. The argumentation becomes trivial if one uses for Majorana
fermions the same formula for charge conjugation as the one
 at the extreme right of (\ref{eq:Cdirac})  for Dirac fermions
$(\psi_M)^c = \gamma^2 (\psi_M)^\ast$, $(\chi_M)^c = \gamma^2(\chi_M)^\ast$.}
.

The Majorana conditions linking $\xi$ and $\eta$  are
\begin{equation}
\xi^\alpha \stackrel{C=\pm 1}{=}
\pm (-i)(\eta^{\dot\alpha})^\ast
\Leftrightarrow
\eta_{\dot\beta}\stackrel{C=\pm 1}{=}\pm (-i)(\xi_\beta)^\ast;
\label{eq:Majocon}
\end{equation}
using formul\ae (\ref{eq:Cland1},\ref{eq:Cland1c})
 for the charge conjugates of Weyl fermions, they also write
\begin{equation}
\xi^\alpha \stackrel{C=\pm 1}{=}\pm  (\xi^\alpha)^c,\quad
\eta_{\dot\beta}\stackrel{C=\pm 1}{=}\pm (\eta_{\dot\beta})^c.
\label{eq:Majocon2}
\end{equation}
A Majorana bi-spinor can accordingly also be written
\footnote{The Majorana spinors $\psi_M^\pm$ and $\chi_M^\pm$ can also be
written
\begin{equation}
\psi_M^\pm = \left(\begin{array}{c} \xi^\alpha \cr \pm
(-i)(\xi^\alpha)^{CP} \end{array}\right), \quad
\chi_M^\pm = \left(\begin{array}{c} \pm(-i) (\eta_{\dot\beta})^{CP} \cr
\eta_{\dot\beta} \end{array}\right);
\label{eq:CCP}
\end{equation}
they involve one Weyl spinor and its $CP$ conjugate (see subsection
\ref{subsection:CPtrans}).}

\begin{equation}
\chi_{M}^\pm = \left(\begin{array}{c}  \pm (-i)
(\eta^{\dot\alpha})^\ast \cr \eta_{\dot\beta} \end{array}\right),
\label{eq:chiMaj}
\end{equation}
which is identical to $\psi^\pm_M$ by the relations (\ref{eq:Majocon}).
By charge conjugation, using (\ref{eq:Cland1}),
 $\psi_M^+ \stackrel{C}{\leftrightarrow}\chi_M^+,
\psi_M^-\stackrel{C}{\leftrightarrow}-\chi^-_M$.

A so-called Majorana mass term writes
\begin{eqnarray}
\overline{\psi_M} \psi_M &\equiv&
\psi_M^\dagger \gamma^0 \psi_M \equiv \pm i\left[ -(\xi^\alpha)^\ast (\xi
_\alpha)^\ast + \xi_\alpha \xi^\alpha \right]
= \pm i\left[ (\xi_\alpha)^\ast (\xi
^\alpha)^\ast + \xi_\alpha \xi^\alpha \right] \cr
or\quad\overline{\psi_M} \gamma^5 \psi_M &\equiv&
\psi_M^\dagger \gamma^0 \gamma^5 \psi_M
\equiv \mp i\left[ (\xi^\alpha)^\ast (\xi
_\alpha)^\ast + \xi_\alpha \xi^\alpha \right]
=\mp i\left[ (-\xi_\alpha)^\ast (\xi
^\alpha)^\ast + \xi_\alpha \xi^\alpha \right].
\label{eq:Majmass}
\end{eqnarray}
Along the same lines, Majorana kinetic terms write
\footnote{One defines as usual $\psi \overleftrightarrow\partial\chi
= \frac12 \big(\psi \partial \chi - (\partial\psi) \chi\big)$.
 For anticommuting
fermions $[\psi,\chi]_+=0$, one has 
 $\psi \overleftrightarrow\partial\chi =\psi
\partial \chi = \chi \partial \psi =\chi \overleftrightarrow\partial\psi $.}
$\overline{\psi_M} \gamma^\mu \overleftrightarrow{p_\mu} \psi_M$
or $\overline{\psi_M} \gamma^\mu\gamma^5 \overleftrightarrow{p_\mu}
\psi_M$; they rewrite in terms of Weyl spinors (using (\ref{eq:kin}))

\vbox{
\begin{eqnarray}
\overline{\psi_M} \gamma^\mu \overleftrightarrow{p_\mu} \psi_M
&=&\psi_M^\dagger \left(\begin{array}{cc}
\overleftrightarrow{(p^0-\vec p.\vec\sigma)} & 0
\cr 0 & \overleftrightarrow{(p^0+\vec p.\vec\sigma)}\end{array}\right)
\psi_M\cr
&&= (\xi^\alpha)^\ast
\overleftrightarrow{(p^0-\vec p.\vec\sigma)} \xi^\beta
+ \big(\pm (-i) (\xi_\alpha)^\ast\big)^\ast
\overleftrightarrow{(p^0+\vec p.\vec\sigma)}
\big(\pm (-i) \xi_\beta^\ast\big)\cr
&&= (\xi^\alpha)^\ast
\overleftrightarrow{(p^0-\vec p.\vec\sigma)} \xi^\beta
+ \xi_\alpha
\overleftrightarrow{(p^0+\vec p.\vec\sigma)} \xi_\beta^\ast,
\label{eq:Majkin1}
\end{eqnarray}
}

and

\vbox{
\begin{eqnarray}
\overline{\psi_M} \gamma^\mu\gamma^5 \overleftrightarrow{p_\mu} \psi_M
&=&\psi_M^\dagger \left(\begin{array}{cc}
\overleftrightarrow{(p^0-\vec p.\vec\sigma)} & 0
\cr 0 & \overleftrightarrow{(p^0+\vec p.\vec\sigma)}\end{array}\right)
\gamma^5\psi_M\cr
&&= (\xi^\alpha)^\ast
\overleftrightarrow{(p^0-\vec p.\vec\sigma)} \xi^\beta
- \big(\pm (-i) (\xi_\alpha)^\ast\big)^\ast
\overleftrightarrow{(p^0+\vec p.\vec\sigma)}
\big(\pm (-i) \xi_\alpha^\ast\big)\cr
&&= (\xi^\alpha)^\ast
\overleftrightarrow{(p^0-\vec p.\vec\sigma)} \xi^\beta
- \xi_\alpha
\overleftrightarrow{(p^0+\vec p.\vec\sigma)} \xi_\beta^\ast.
\label{eq:Majkin2}
\end{eqnarray}
}

A Dirac fermion can always be written as the sum of two Majorana's (the first
has $C=+1$ and the second $C=-1$):
$\left(\begin{array}{c} \xi^\alpha \cr \eta_{\dot\beta}\end{array}\right) =
\frac{1}{2}\left[
\left(\begin{array}{c} \xi^\alpha - i(\eta^{\dot\alpha})^\ast \cr
-i\xi_\beta^\ast + \eta_{\dot\beta}\end{array}\right)
+ \left(\begin{array}{c} \xi^\alpha +i(\eta^{\dot\alpha})^\ast \cr
i\xi_\beta^\ast + \eta_{\dot\beta}\end{array}\right)\right]$.

While a Dirac fermion $\pm$ its charge conjugate is always a Majorana fermion
($C=\pm1$),
any Majorana fermion ({\em i.e.} a general bi-spinor which is a $C$
eigenstate) cannot be uniquely written as the sum of a given Dirac fermion
$\pm$ its charge conjugate: this decomposition is not unique.
Suppose indeed that, for example, a $C=+1$ Majorana fermion is written like
the sum of a Dirac fermion + its charge conjugate
$\left(\begin{array}{c} \theta^\alpha \cr -i \theta_\beta^\ast
\end{array}\right)
= \left(\begin{array}{c} \xi^\alpha -i(\eta^{\dot\alpha})^\ast \cr
\eta_{\dot\beta} -i \xi_\beta^\ast \end{array}\right)$. Since the two
corresponding equations are not independent, $\xi$ and $\eta$ cannot be fixed,
but only the combination $\xi^\alpha -i(\eta^{\dot\alpha})^\ast \sim
\xi^\alpha -i\eta^\alpha$. So, infinitely many different
 Dirac fermions can be used for this purpose.

A Majorana fermion can always be written  as the sum of a left fermion
$\pm$
its charge conjugate, or the sum of a right fermion $\pm$ its charge conjugate.
Let us demonstrate the first case only, since the second goes exactly along
the same lines
\begin{eqnarray}
\psi^\pm_M &=& \left(\begin{array}{c} \xi^\alpha \cr
\pm(-i)\xi_\beta^\ast\end{array}\right)
= \left(\begin{array}{c} \xi^\alpha \cr 0\end{array}\right)
+\left(\begin{array}{c} 0 \cr \pm(-i)\xi_\beta^\ast\end{array}\right)
= \psi_L \pm \gamma^2 \psi_L^\ast
= \psi_L \pm (\psi_L)^c,\cr
\psi_L &=& \left(\begin{array}{c} \xi^\alpha \cr 0\end{array}\right)
=\frac{1+\gamma^5}{2}\psi_D.
\end{eqnarray}

Majorana fermions have $PC\,parity=\pm i$.
For example, $PC.\left(\begin{array}{c} \xi^\alpha \cr (\eta_{\dot\beta})^c
\end{array}\right) =
\left(\begin{array}{c} \xi_\alpha^\ast \cr i\xi ^\beta
\end{array}\right) 
=i\gamma^0 \left(\begin{array}{c} \xi^\alpha \cr (\eta_{\dot\beta})^c
\end{array}\right)$.
They are {\em not} $PC$ eigenstates (an extra $\gamma^0$ comes into play
in the definition of $PC$-{\em parity}).

\section{Invariance}
\label{section:invar}

\subsection{Wightman's point of view \cite{Wightman}}
\label{subsection:invar1}

The invariance of a ``theory'' is  expressed by the invariance of the
vacuum and the invariance of all $n$-point functions; $\cal O$ is then a
product of fields at different space-time points and ($\hat{\cal O}$
being the transformed of $\cal O$)
\begin{equation}
|\ 0> = |\ \hat 0>,
<0\ |\ {\cal O}\ |\ 0> = <0\ |\ \hat{\cal O}\ |\ 0>.
\label{eq:Winvar}
\end{equation}
$\ast$ in the case of a unitary transformation ${\cal U}$,
\begin{equation}
<0\ |\ {\cal O}\ |\ 0> \stackrel{sym}{=} <0\ |\ {\cal O}^U\ |\ 0>
\stackrel{vacuum\ inv}{=} <0^U\ |\ {\cal O}^U\ |\ 0^U>,\
{\cal O}^U = {\cal U}^{-1}{\cal O}{\cal U};
\end{equation}
taking the example of parity and if ${\cal O} = \phi_1(x_1)\phi_2(x_2)\ldots
\phi_n(x_n)$, one has

${\cal O}^P = P^{-1}{\cal O}P =
\phi_1(t_1,-\vec x_1) \phi_2(t_2,-\vec x_2) \ldots \phi_n(t_n,-\vec x_n)$,
such that parity invariance writes
\begin{equation}
<0\ |\ \phi_1(x_1) \phi_2(x_2)\ldots \phi_n(x_n)\ |\ 0> =
<0\ |\ \phi_1(t_1,-\vec x_1) \phi_2(t_2,-\vec x_2) \ldots \phi_n(t_n,-\vec
x_n)\ |\ 0>.
\end{equation}

$\ast$ in the case of a antiunitary transformation ${\cal A}$,
\begin{eqnarray}
<0\ |\ {\cal O}\ |\ 0> &\stackrel{sym}{=}& <0\ |\ {\cal O}^A\ |\ 0>
= <0^A\ |\ {\cal O}^A\ |\ 0^A>;\cr
{\cal O}^A &=& ({\cal A}^{-1}{\cal O}{\cal A})^\dagger \Rightarrow \cr
<0\ |\ {\cal O}\ |\ 0> &\stackrel{sym}{=}&
= <0\ |\ ({\cal A}^{-1}{\cal O}{\cal A})^\dagger\ |\ 0>
=<0\ |\ {\cal A}^{-1}{\cal O}{\cal A}\ |\ 0>^\ast;\cr
&&
\label{eq:antitrans}
\end{eqnarray}
taking the example of $\Theta = PCT$, with
 ${\cal O} = \phi_1(x_1)\phi_2(x_2)\ldots \phi_n(x_n)$, one has

${\cal O}^\Theta = (\Theta^{-1}{\cal O}\Theta)^\dagger =
(\Theta^{-1}\phi_n \Theta)^\dagger \ldots  (\Theta^{-1}\phi_2
\Theta)^\dagger (\Theta^{-1}\phi_n \Theta)^\dagger
= \phi_n^\Theta \ldots \phi_2^\Theta \phi_1^\Theta$.

For fermions \cite{Wightman}
\begin{equation}
\phi(x)^\Theta \equiv \pm \phi(-x) =(\Theta^{-1} \phi(x) \Theta)^\dagger,
\end{equation}
 such that $PCT$ invariance
expresses as (of course the sign is unique and must be precisely
determined)

\vbox{
\begin{eqnarray}
<0\ |\  \phi_1(x_1)\phi_2(x_2)\ldots \phi_n(x_n)\ |\ 0> &\stackrel{sym}
{=}& \pm <0\ |\ \phi_n(-x_n)\ldots \phi_2(-x_2)\phi_1(-x_1)\ |\ 0>\cr
&&\hskip -3cm = \pm <0\ |\ \phi_1^\ast(-x_1)\phi_2^\ast(-x_2)\ldots \phi_n^\ast(-x_n)\ |\
0>^\ast\cr
 &&\hskip -3cm =\pm <0\ |\
(\Theta^{-1}\phi_1(x_1)\Theta)(\Theta^{-1}\phi_2(x_2)\Theta)
\ldots (\Theta^{-1}\phi_n(x_n)\Theta)\ |\ 0>^\ast.\cr
&&
\end{eqnarray}
}

It is enough to change $x_i \to -x_i$ and to read all Green functions from
right to left instead of reading them from left to right (like Pauli).

For a general antiunitary transformation $\cal A$, 
the last line of (\ref{eq:antitrans})
expressing the invariance also reads,
since the vacuum is supposed to be invariant by ${\cal A}^{-1}$ as well as
by $\cal A$:
\begin{eqnarray}
&& <0\ |\ {\cal O}\ |\ 0>\equiv <0\ |\ {\cal O}\,0>\cr
 &&=<{\cal A}^{-1} 0\ |\ ({\cal A}^{-1}{\cal O}{\cal A})^\dagger\ |\ {\cal
A}^{-1}0>
= <{\cal A}^{-1}0\ |\ ({\cal A}^{-1}{\cal O}{\cal A})\ |\ {\cal A}^{-1}0>^\ast =
<{\cal A}^{-1} 0\ |\ {\cal A}^{-1} ({\cal O}\,0)>^\ast;\cr
&&
\label{eq:atrans2}
\end{eqnarray}
requesting that, for any $\phi$, $<\phi\ |\ {\cal O}\ |\ \phi>
= <\phi\ |\ (A^{-1}{\cal O}A)^\dagger\ |\ \phi>$ would be much stronger a
condition.

\bigskip

Wightman's expression of the invariance is  weaker than requesting
${\cal O} = \hat{\cal O}$, since it occurs only for VEV's and not when
sandwiched between any state $\phi$.

\subsection{The condition ${\cal O} = \hat{\cal O}$}

It is often used to express the invariance of a theory with
(Lagrangian or) Hamiltonian $\cal O$ by the transformation under
consideration:

* For unitary transformations, this condition is equivalent to
\begin{equation}
{\cal O} ={\cal U}^{-1}{\cal O}{\cal U}
\Leftrightarrow [{\cal U},{\cal O}]=0;
\label{eq:invuni}
\end{equation}
* For antiunitary transformations it yields
(we use the property  that, for unitary as well as
for antiunitary operators ${\cal U}^{-1} = {\cal U}^\dagger$ and ${\cal
A}^{-1} = {\cal A}^\dagger$, see Footnote \ref{footnote:Weinberg} and
Appendix \ref{section:adjointanti})
\begin{equation}
{\cal O} =({\cal A}^{-1}{\cal O}{\cal A})^\dagger =
{\cal A}^{-1}{\cal O}^\dagger {\cal A} \Leftrightarrow {\cal A}{\cal O} =
{\cal O}^\dagger {\cal A}.
\label{eq:invantiuni}
\end{equation}
Note that this is similar (apart from the exchange $\Theta \leftrightarrow
\Theta^{-1}$) to the condition proposed in \cite{Lee} (p.322)  as the
``$PCT$'' theorem for any Lagrangian density ${\cal L}(x)$ considered as a
hermitian {\em operator}
\begin{equation}
\Theta {\cal L}(x) \Theta^{-1} = {\cal L}^\dagger(-x).
\label{eq:CPTLee}
\end{equation}
So, that the Hamiltonian commutes with the symmetry
transformation can eventually be accepted when this transformation is unitary
(and we have already mentioned that this statement is stronger that
Wightman's expression for invariance); however, when the
transformation is antiunitary, one must be more careful.

Requesting that the transformed states should satisfy the same equations
as the original ones is only true for unitary transformations. It is not
in the case of antiunitary operations like $T$ (or $PCT$)
since a time reversed  fermion does not satisfy the same equation as the
original fermion but the time-reversed equation.

\subsection{Hamiltonian - Lagrangian.}

\subsubsection{The case of a unitary transformation}
\label{subsub:transunit}

$\bullet$\ {\bf Invariance of the Hamiltonian:}

In Quantum Mechanics, a system is said to be invariant by a unitary
transformation ${\cal U}$ if the transformed of the eigenstates of the Hamiltonian
$H$ have the same energies as the original states
\begin{equation}
H \psi = E\psi\ and \ H {\cal U}\cdot \psi = E {\cal U}\cdot \psi;
\label{eq:Hinv1}
\end{equation}
since ${\cal U}$ is unitary, it is in particular linear, such that $E{\cal
U}\cdot \psi = {\cal U}\cdot 
E\psi = {\cal U}\cdot H\psi$; this is why the invariance of the theory is commonly
 expressed by
\begin{equation}
H = {\cal U}^{-1}H{\cal U} \Leftrightarrow [{\cal U},H]=0.
\label{eq:invarunit}
\end{equation}
Defining, according to Wightman, the transformed $\hat H$ of the
Hamiltonian $H$ by $\hat H = {\cal U}^{-1}H{\cal U}$, we see the the invariance condition
(\ref{eq:invarunit}) also rewrites $\hat H=H$. No special condition of
reality is required for $E$.

$\bullet$\ {\bf Invariance of the Lagrangian:}

The Lagrangian approach is often more convenient in Quantum Field Theory;
it determines the (classical) equations of motion, and also the
perturbative expansion.
The Lagrangian density ${\cal L}(x)$ is written $<\Psi(x)\ |\ L(x)\ |\
\Psi(x)>$,
where $L$ is an operator and $\Psi(x)$ is a ``vector'' of different  fields.

A reasonable definition for the invariance of the theory if that the
transformed ${\cal U}\Psi$ of $\Psi$ satisfies the same equation as $\Psi$; since
${\cal L}(x)$ and $e^{i\alpha}{\cal L}(x)$ will provide the same
(classical) dynamics, one expresses this invariance by
\begin{equation}
<{\cal U}\cdot \Psi(x)\ |\ L(x)\ |\ {\cal U}\cdot \Psi(x)> = e^{i\alpha}<\Psi(x)\ |\ L(x)\ |\ \Psi(x)>
=e^{i\alpha}{\cal L}(x).
\label{eq:Uinv}
\end{equation}
Due to the unitarity of ${\cal U}$, this is equivalent to
$<\Psi(x)\ |\ {\cal U}^{-1} L(x){\cal U}\ |\ \Psi(x)> = e^{i\alpha}
<\Psi(x)\ |\ L(x)\ |\ \Psi(x)>$
or, owing to the fact that $\Psi$ can be anything
\begin{equation}
L{\cal U}=e^{i\alpha}{\cal U}L.
\label{eq:Linvuni}
\end{equation}
If one applies this rule to a mass term, and consider the mass (scalar) 
as an operator, the unitarity of $\cal U$ entails that a scalar as well as
the associated operator should  stay unchanged. This
leaves only the possibility $\alpha =0$. The condition (\ref{eq:Linvuni}) 
reduces accordingly to the vanishing of the commutator $[L,{\cal U}]$.
Wightman's definition (\ref{eq:Wtrans}) of the transformed
$\hat L = {\cal U}^{-1}L{\cal U}$ of the operator $L$ makes this condition equivalent
to $\hat L =  L$. No condition of reality (hermiticity) is required on $L$.

\subsubsection{The case of antiunitary transformations}
\label{subsub:transanti}

The situation is more tricky, since, in particular, the states transformed
by a antiunitary transformation (for example $T$) do not satisfy the same
classical equations as the original states (in the case of $T$, they
satisfy the time-reversed equations).

This why it is more convenient to work with each bilinear present in the
Lagrangian or Hamiltonian, which we write for example $< \phi\ |\ {\cal O}\
|\ \chi>$. $\phi, \xi$ can be fermions or bosons, $\cal O$ a scalar, a
derivative operator \ldots.
Taking the example of $PCT$, this bilinear transforms into
$< \Theta \phi\ |\ {\cal O}\ |\ \Theta \chi>
\stackrel{(\ref{eq:antiunit5})}{=} <\chi\ |\ \hat{\cal O}\ |\
\phi> =\break <\chi \ |\ (\Theta^{-1}{\cal O} \Theta)^\dagger\ |\ \phi>$.

\bigskip

{\bf Application: Dirac and Majorana mass terms}

\smallskip

$\bullet$\ \underline{{\it Problems with a classical fermionic Lagrangian:}}
\medskip

In view of all possible terms compatible with Lorentz invariance, we work
in a basis which can accommodate, for example,
both a Dirac fermion and its antiparticle. Accordingly,
For a single Dirac fermion (and its antiparticle), we introduce the
4-vector of Weyl fermions
\begin{equation}
\psi
=\left(\begin{array}{c} n_L \cr n_R \end{array}\right)
= \left(\begin{array}{c}
\xi^\alpha \cr (\xi^\beta)^c \cr (\eta_{\dot\gamma})^c \cr \eta_{\dot\delta}
\end{array}\right)
\equiv
\left(\begin{array}{c}
\xi^\alpha \cr -i(\eta^{\dot\beta})^\ast \cr -i(\xi_{\gamma})^\ast \cr
\eta_{\dot\delta} \end{array}\right)
\stackrel{Lorentz}{\sim} \left(\begin{array}{c} \xi^\alpha \cr \eta^\beta \cr \xi_{\dot\gamma}
\cr \eta_{\dot\delta} \end{array}\right),
\label{eq:basis}
\end{equation}
where $\stackrel{Lorentz}{\sim}$ means ``transforms like (by Lorentz)''.

 Let us study the
transform by $PCT$ of a Dirac-type mass term
$m_D \xi^{\alpha\ast}(x) \eta_{\dot\alpha}(x)=
<\xi^\alpha(x)\ |\ m_D\ |\ \eta_{\dot\alpha}(x)>$
and of a Majorana-type mass term
$m_M \xi^{\alpha\ast}(x) (\eta_{\dot\alpha})^c(x) 
= <\xi^\alpha(x)\ |\ m_M\ |\ (\eta_{\dot\alpha})^c(x)>$.

*\ $m_D$ and $m_M$ we first
 consider as operators sandwiched between fermionic grassmanian functions.
The two mass terms transform, respectively, into
$<\Theta \xi^\alpha(x)\ |\ m_D\ |\ \Theta \eta_{\dot\alpha}(x)>$
and $<\Theta \xi^\alpha(x)\ |\ m_M\ |\ \Theta (\eta_{\dot\alpha})^c(x)>$.
We now use (\ref{eq:antiunit5}), which transforms these two expressions into
$<\eta_{\dot\alpha}\ |\ m_D^\Theta\ |\ \xi^\alpha>$ and
$<(\eta_{\dot\alpha})^c\ |\ m_M^\Theta \ |\ \xi^\alpha>$.
Since $\Theta$ is antilinear, $\Theta^{-1} m \Theta = m^\ast \Rightarrow
m^\Theta \equiv (\Theta^{-1} m \Theta)^\dagger = m$.
So the two mass terms transform,
respectively, into $m_D <\eta_{\dot\alpha}\ |\ \xi^\alpha> \equiv
m_D \eta_{\dot\alpha}^\ast \xi^\alpha$
 and $m_M <\eta_{\dot\alpha}^c\ |\ \xi^\alpha> \equiv m_M
(\eta_{\dot\alpha}^c)^\ast \xi^\alpha$. Notice that
$\eta_{\dot\alpha}^\ast \xi^\alpha$ is (using anticommutation) $(-)$ the
complex conjugate of $\xi^{\alpha\ast}\eta_{\dot\alpha}$ and likewise,
that $(\eta_{\dot\alpha}^c)^\ast \xi^\alpha$ is $(-)$ the complex conjugate
of $\xi^{\alpha\ast} \eta_{\dot\alpha}^c$.

The Lagrangian density also a priori involves Dirac and Majorana mass terms
$\mu_D \eta_{\dot\alpha}^\ast \xi^\alpha$ and
$\mu_M (\eta_{\dot\alpha}^c)^\ast \xi^\alpha$, such that $PCT$ invariance
requires $m_D = \mu_D$ and $m_M=\mu_M$
\footnote{If the Lagrangian (Hamiltonian) is furthermore real, it should
match its complex conjugate (see Appendix \ref{section:cchermit}).
The c.c. of the Dirac mass terms are
$m_D^\ast \xi^{\alpha} \eta_{\dot\alpha}^\ast + \mu_D^\ast
\eta_{\dot\alpha} \xi^{\alpha\ast}   \stackrel{anticom}{=} - m_D^\ast 
\eta_{\dot\alpha}^\ast \xi^{\alpha}
 -\mu_D^\ast\xi^{\alpha\ast} \eta_{\dot\alpha}$
and the c.c. of the Majorana mass term are $m_M^\ast \xi^{\alpha}
(\eta_{\dot\alpha}^c)^{\ast} + \mu_M^\ast (\eta_{\dot\alpha}^c)
\xi^{\alpha\ast} \stackrel{anticom}{=}  -m_M^\ast (\eta_{\dot\alpha}^c)^{\ast}\xi^{\alpha}
 -\mu_M^\ast \xi^{\alpha\ast}  (\eta_{\dot\alpha}^c)$. Using
(\ref{eq:Cland1}) to
replace $\eta_{\dot\alpha}^c$ by $(-i) \xi_{\alpha}^\ast$, the reality of the
Lagrangian is seen to require $m_D = -\mu_D^\ast$ and $m_M = -\mu_M^\ast$.

So, combining the two, we see that a real and $PCT$ invariant (classical)
Lagrangian should satisfy $m_D=\mu_D\ imaginary$ and $m_M=\mu_M\
imaginary$.}.

\smallskip

*\ If we instead consider that $m\phi^\ast \chi \stackrel{PCT}{\rightarrow}
m (\Theta \phi^\ast) \Theta \chi$ we obtain, using
(\ref{eq:TCPland1b}) and (\ref{eq:TCPland2b}),
 that the Dirac mass term
transforms into $m_d (-i\xi^{\alpha\ast}) (-i \eta_{\dot\alpha})$, that is,
it changes sign by $PCT$. The Majorana mass term transforms into
$m_M (-i \xi^{\alpha\ast})\Theta(-i \xi_\alpha^\ast)
\stackrel{antilin}{=} m_M (-i \xi^{\alpha\ast}) (+i) \Theta \xi_\alpha^\ast
= (-i \xi_\alpha^\ast) (+i) (-i \xi_\alpha^\ast)
= -i \xi^{\alpha\ast}\xi_\alpha^\ast$, that is, unlike the Dirac mass term,
 the Majorana mass term does not change sign.
This alternative would in particular exclude the simultaneous presence of
Dirac and Majorana mass terms (necessary for the see-saw mechanism).

*\ Conclusion: antiunitary transformations of a classical fermionic
Lagrangian are ambiguous and can lead to contradictory statements.
Defining a classical fermionic Lagrangian is most probably itself
problematic
\footnote{Let us also mention the arbitrariness that results from adding to
a mass matrix any vanishing anticommutator.}.

\smallskip

$\bullet$\ \underline{{\it Quantum (operator) Lagrangian}}

Dirac and Majorana mass terms write, respectively
$[\xi^\alpha]^\dagger [m_D] [\eta_{\dot\alpha}]$ and
$[\xi^\alpha]^\dagger [m_M] [\eta_{\dot\alpha}^c] 
\stackrel{(\ref{eq:psiopC})}{=}
 [\xi^\alpha]^\dagger  [m_M] (-i) [\xi_\alpha]^\dagger$.

Using (\ref{eq:antiumit4}), one gets
$([\xi^\alpha]^\dagger [m_D] [\eta_{\dot\alpha}])^\Theta = 
[\eta_{\dot\alpha}]^\Theta [m_D]^\Theta ([\xi^\alpha]^\dagger)^\Theta
=[\eta_{\dot\alpha}]^\Theta [m_D]^\Theta ([\xi^\alpha]^\Theta)^\dagger
= -i[\eta_{\dot\alpha}] [m_D] (-i)[\xi^\alpha]^\dagger$, such that, using the
anticommutation of fermionic operators, the Dirac
mass term transforms by $\Theta$ into itself.\newline
As far as the Majorana mass term is concerned, it transforms into
$([\xi^\alpha]^\dagger [m_M] [\eta_{\dot\alpha}^c])^\Theta
= ([\eta_{\dot\alpha}^c])^\Theta [m_M]^\Theta
([\xi^\alpha]^\dagger)^\Theta
= (-i[\xi_\alpha]^\dagger)^\Theta [m_M]^\Theta
([\xi^\alpha]^\dagger)^\Theta$. 
One uses again (\ref{eq:antiumit4})
to evaluate $(-i[\xi_\alpha]^\dagger)^\Theta = ([\xi_\alpha]^\dagger)^\Theta 
(-i)^\Theta = (-i)[\xi_\alpha]^\dagger (-i) = -[\xi_\alpha]^\dagger$.
So, finally, the Majorana mass term transforms into
$-[\xi_\alpha]^\dagger m_M (-i)[\xi^\alpha]^\dagger \stackrel {anticom}{=}
-i [\xi^\alpha]^\dagger m_M [\xi_\alpha]^\dagger$, that is, like the Dirac
mass term, into itself.\newline
 The same conclusions are obtained in the propagator formalism.

\section{The fermionic propagator and discrete symmetries (1 fermion + its
antifermion)}

The fermionic propagator $\Delta(x)$
 is a matrix with a Lorentz tensorial structure,
the matrix elements of which are the vacuum expectation values
of $\cal T$-products of two fermionic operators:
\begin{equation}
{\cal T} \psi(x) \chi(y) = \theta (x^0-y^0) \psi(x)\chi(y) -\theta (y^0-x^0)
\chi(y)\psi(x);
\label{eq:Tprod}
\end{equation}
the Lorentz indices of the two operators yield the tensorial structure of
the matrix elements.

If, for example, one works in the fermionic basis
$(\psi_1,\psi_2,\psi_3,\psi_4)$, and if $\alpha,\beta \ldots $ denote their
Lorentz indices, the propagator is a $4 \times 4$ matrix
$\Delta(x)$ such that
\begin{equation}
\Delta_{ij}^{\alpha\beta}(x) = <\psi_i^\alpha\ |\Delta(x)|\ \psi_j^\beta>=
<0\ |{\cal T} (\psi_i)^\alpha(\frac{x}{2}) (\psi_j^\dagger)^\beta(-\frac{x}{2})|\ 0>.
\end{equation}
Supposing
\begin{equation}
<\psi_i^\alpha\ |\ \psi_j^\beta> = \delta_{ij}\delta^{\alpha\beta},
\end{equation}
we shall also use the notation, 

\vbox{
\begin{eqnarray}
\Delta(x) &=&
\sum_{i,j} |\ \psi_i^\alpha>\Delta_{ij}^{\alpha\beta}(x)<\psi_j^\beta\ |\cr
&=& \left(\begin{array}{cccc}|\ \psi_1^\alpha> & |\ \psi_2^\alpha> & |\
\psi_3^\alpha> & |\ \psi_4^\alpha>\end{array}\right)
\Delta_{ij}^{\alpha\beta}(x)
\left(\begin{array}{c}
   <\psi_1^\beta\ |   \cr <\psi_2^\beta\ |  \cr <\psi_3^\beta\ |  \cr
<\psi_4^\beta\ | \end{array}\right);
\end{eqnarray}
\label{eq:notaprop}
}

since one indeed finds $<\psi_i^\alpha\ |\Delta(x)|\ \psi_j^\beta> =
\Delta_{ij}^{\alpha\beta}(x)$.

We will work hereafter in the basis (\ref{eq:basis}), which includes
enough  degrees of freedom to describe a (Dirac) fermion + its antifermion.
The corresponding fermionic propagator is then a $4 \times 4$ matrix which
 involves the following types of ${\cal T}$-products
\footnote{For the Lagrangian, the equivalent would be to consider all
possible quadratic terms compatible with Lorentz invariance. Dirac as well
as Majorana mass terms are allowed, and for, kinetic terms, diagonal ones,
for example $\xi^{\alpha\dagger}(p^0 - \vec p.\vec
\sigma)_{\alpha\beta}\,\xi^\beta$ as well as non-diagonal ones, for example
$(\eta_{\dot\alpha})^{c\dagger} (p^0 + \vec
p.\vec\sigma)_{\alpha\beta}\,\eta_{\dot\beta} \equiv
i\xi_\alpha (p^0 + \vec
p.\vec\sigma)_{\alpha\beta}\,\eta_{\dot\beta}$.}

$\ast$ {mass-like propagators}:
\medskip

\hskip 2cm $<0\ |{\cal T} \xi^\alpha(x) (\eta_{\dot\beta})^\dagger(-x)|\ 0>$ and
$<0\ |{\cal T} (\xi^\alpha)^{c}(x) ((\eta_{\dot\beta})^{c})^\dagger(-x)|\ 0>$
(Dirac-like),

\hskip 2cm $<0\ |{\cal T} (\eta_{\dot\alpha})^{c}(x) ((\xi^\beta)^c)^\dagger(-x)|\ 0>$
and
$<0\ |{\cal T} \eta_{\dot\alpha}(x) (\xi^\beta)^\dagger(-x)|\ 0>$
(Dirac-like),

\hskip 2cm $<0\ |{\cal T} \xi^\alpha(x) ((\eta_{\dot\beta})^c)^\dagger(-x)|\ 0>$,
$<0\ |{\cal T} (\xi^\alpha)^{c}(x) (\eta_{\dot\beta})^\dagger(-x)|\ 0>$
(Majorana-like),

\hskip 2cm$<0\ |{\cal T} (\eta_{\dot\alpha})^{c}(x) (\xi^\beta)^\dagger(-x)|\ 0>$,
$<0\ |{\cal T} \eta_{\dot\alpha}(x) ((\xi^\beta)^c)^\dagger(-x)|\ 0>$
(Majorana-like);

\smallskip

$\ast$ {kinetic-like propagators}:
\medskip

\vbox{
\hskip 2cm $<0\ |{\cal T} \xi^\alpha(x) (\xi^\beta)^\dagger(-x)|\ 0>$ and
$<0\ |{\cal T} (\xi^\alpha)^{c}(x) ((\xi^\beta)^c)^\dagger(-x)|\ 0>$ (diagonal),

\hskip 2cm $<0\ |{\cal T} (\eta_{\dot\alpha})^{c}(x)
((\eta_{\dot\beta})^c)^\dagger(-x)|\ 0>$ and
 $<0\ |{\cal T} \eta_{\dot\alpha}(x)
(\eta_{\dot\beta})^\dagger(-x)|\ 0>$ (diagonal),

\hskip 2cm $<0\ |{\cal T} \xi^\alpha(x) ((\xi^\beta)^c)^\dagger(-x)|\ 0>$ and
$<0\ |{\cal T} (\xi^\alpha)^{c}(x) (\xi^\beta)^\dagger(-x)|\ 0>$
(non-diagonal),

\hskip 2cm $<0\ |{\cal T} (\eta_{\dot\alpha})^{c}(x) (\eta_{\dot\beta})^\dagger(-x)|\ 0>$ and
$<0\ |{\cal T} \eta_{\dot\alpha}(x)((\eta_{\dot\beta})^c)^\dagger(-x)|\ 0>$
(non-diagonal).
}

\medskip

Because of electric charge conservation, some of the mixed
propagators (Majorana mass terms, non-diagonal kinetic terms)
  will only occur for neutral fermions.

Any propagator is a non-local functional of  two fields, which are
evaluated at two different space-time points; a consequence is that, unlike
for the Lagrangian, which is a local functional of the fields, one cannot
implement constraints coming from the anticommutation of fermions.
Likewise, a propagator has no hermiticity (or reality) property, and no
corresponding constraint exist
\footnote{Only the spectral function has positivity properties.}
.
So, the only constraints that can be cast on the propagator come from
discrete symmetries and their combinations: $C$, $CP$, $PCT$.
The mass eigenstates, which are determined from the propagator
are accordingly expected to be less constrained than the
eigenstates of any quadratic Lagrangian
\footnote{and any mass matrix, which can only be eventually introduced
in a linear approximation to the inverse propagator in the vicinity of one
of its poles \cite{Novikov}.}
.

\subsection{$\boldsymbol{PCT}$ constraints}
\label{subsection:TCPprop}

All demonstrations proceed along the following steps.

Suppose that we want to deduce $PCT$ constraints for
$<0\ |{\cal T} \psi(x) \chi^\dagger(-x)|\ 0>$. The information that we have
from (\ref{eq:TCPop}) is:
there exist $\phi$ and $\omega$ such that
$\psi(x) = \Theta \phi^\dagger(-x)\Theta^{-1}$, $\chi^\dagger(-x) =
\Theta\, \omega(x)\Theta^{-1}$
\footnote{For example, from (\ref{eq:TCPop}), one gets
 $\xi^\alpha= \Theta (-i(\xi^\alpha)^\dagger) \Theta^{-1}$.}
,
 the vacuum is supposed to be invariant $|\ 0> = |\ \Theta\; 0>$,
 and $\Theta$ is antiunitary, which entails (\ref{eq:unianti2})
\footnote{$\Theta$, though antiunitary, does not act on the $\theta$
functions of the $\cal T$-product because they  are real.}
.
We have accordingly

$<0\ |{\cal T} \psi(x) \chi^\dagger(-x)|\ 0> =
<0\ | {\cal T} \Theta \phi^\dagger(-x)\Theta^{-1} \Theta\, \omega(x)
\Theta^{-1}|\ 0>
\newline
\stackrel{invariance\ of\ the\ vacuum}{=}
<\Theta\; 0\ | {\cal T} \Theta \phi^\dagger(-x)\Theta^{-1} \Theta
\, \omega(x)\Theta^{-1}|\ \Theta\; 0> 
=<\Theta\; 0\ | {\cal T} \Theta \phi^\dagger(-x) \omega(x)\Theta^{-1}|\
\Theta\; 0>$\newline
$\stackrel{(\ref{eq:unianti2})}{=}
<0\ |\theta(t) \omega^\dagger(x) \phi(-x) -\theta(-t)
\phi(-x)\omega^\dagger(x)|\ 0>
 =- <0\ |{\cal T} \phi(-x) \omega^\dagger(x)|\ 0>$.

\subsubsection{Constraints on mass-like terms}
\label{subsub:sumcons}
%
\begin{eqnarray}
\ast\ Majorana-like\
<0\ |{\cal T} \xi^\alpha(x) ((\eta_{\dot\beta})^c)^\dagger(-x)|\ 0>
&=&
<0\ |{\cal T} \xi^\alpha(-x) ((\eta_{\dot\beta})^c)^\dagger(x)|\ 0> \cr
&=&
 - <0\ |{\cal T} ((\eta_{\dot\beta})^c)^\dagger(x) \xi^\alpha(-x)|\ 0>;\cr
\ast\ Majorana-like\
<0\ |{\cal T} (\eta_{\dot\alpha})^{c}(x) (\xi^\beta)^\dagger(-x)|\ 0>
&=&
<0\ |{\cal T} (\eta_{\dot\alpha})^{c}(-x) (\xi^\beta)^\dagger(x)|\ 0>\cr
&=& - <0\ |{\cal T} (\xi^\beta)^\dagger(x) (\eta_{\dot\alpha})^{c}(-x) |\ 0>
;\cr
\ast\ Majorana-like\
<0\ |{\cal T} (\xi^\alpha)^{c}(x) (\eta_{\dot\beta})^\dagger(-x)|\ 0>
&=&
<0\ |{\cal T} (\xi^\alpha)^{c}(-x) (\eta_{\dot\beta})^\dagger(x)|\ 0>\cr
&=& - <0\ |{\cal T} (\eta_{\dot\beta})^\dagger(x) (\xi^\alpha)^{c}(-x)|\ 0>
;\cr
\ast\ Majorana-like\
<0\ |{\cal T} \eta_{\dot\alpha}(x) ((\xi^\beta)^c)^\dagger(-x)|\ 0>
&=&
<0\ |{\cal T} \eta_{\dot\alpha}(-x) ((\xi^\beta)^c)^\dagger(x)|\ 0>\cr
&=& - <0\ |{\cal T} ((\xi^\beta)^c)^\dagger(x)\eta_{\dot\alpha}(-x)|\ 0>
;\cr
&& \cr
\ast\ Dirac-like\
<0\ |{\cal T} \xi^\alpha(x) (\eta_{\dot\beta})^\dagger(-x)|\ 0>
&=&
<0\ |{\cal T} \xi^\alpha(-x) (\eta_{\dot\beta})^\dagger(x)|\ 0>\cr
&=& - <0\ |{\cal T} (\eta_{\dot\beta})^\dagger(x) \xi^\alpha(-x) |\ 0>:\cr
\ast\ Dirac-like\
<0\ |{\cal T} \eta_{\dot\alpha}(x) (\xi^\beta)^\dagger(-x)|\ 0>
&=&
<0\ |{\cal T} \eta_{\dot\alpha}(-x) (\xi^\beta)^\dagger(x)|\ 0>\cr
&=& - <0\ |{\cal T} (\xi^\beta)^\dagger(x)\eta_{\dot\alpha}(-x) |\ 0>;\cr
\ast\ Dirac-like\
<0\ |{\cal T} (\xi^\alpha)^{c}(x) ((\eta_{\dot\beta})^{c})^\dagger(-x)|\ 0>
&=&
<0\ |{\cal T} (\xi^\alpha)^{c}(-x) ((\eta_{\dot\beta})^{c})^\dagger(x)|\ 0>\cr
&=& - <0\ |{\cal T} ((\eta_{\dot\beta})^{c})^\dagger(x) (\xi^\alpha)^{c}(-x) |\ 0>;\cr
\ast\ Dirac-like\
<0\ |{\cal T} (\eta_{\dot\alpha})^c(x) ((\xi^\beta)^c)^\dagger(-x)|\ 0>
&=&
<0\ |{\cal T} (\eta_{\dot\alpha})^c(-x) ((\xi^\beta)^c)^\dagger(x)|\ 0>\cr
&=& - <0\ |{\cal T} ((\xi^\beta)^c)^\dagger(x)(\eta_{\dot\alpha})^c(-x) |\ 0>.\cr
&&
\label{eq:cons1}
\end{eqnarray}
We give the demonstration of the first (Majorana-like) line of
(\ref{eq:cons1}).

$<0\ |{\cal T} \xi^\alpha(x) ((\eta_{\dot\beta})^c)^\dagger(-x)|\ 0>
= <0\ |{\cal T} \xi^\alpha(x) i\xi_\beta(-x)|\ 0>
= i<0\ |{\cal T} \xi^\alpha(x) \xi_\beta(-x)|\ 0>\newline
= i<0\ |{\cal T} \Theta (-i(\xi^\alpha)^\dagger(-x))\Theta^{-1}
\Theta (-i(\xi_\beta)^\dagger)(x) \Theta^{-1} |\ 0>\newline
\stackrel{invariance\ of\ the\ vacuum}{=}   i<\Theta\,0\ |{\cal T} \Theta (-i(\xi^\alpha)^\dagger)(-x)\Theta^{-1}
\Theta (-i(\xi_\beta)^\dagger)(x) \Theta^{-1} |\ \Theta\,0>\newline
=  i<\Theta\,0\ |{\cal T} \Theta (-i(\xi^\alpha)^\dagger)(-x)
(-i(\xi_\beta)^\dagger)(x) \Theta^{-1} |\ \Theta\,0>\newline
=  -i<\Theta\,0\ |{\cal T} \Theta (\xi^\alpha)^\dagger(-x)
(\xi_\beta)^\dagger(x) \Theta^{-1} |\ \Theta\,0>\newline
\stackrel{antiunitarity (\ref{eq:unianti2})}{=}
 -i<0\ |\theta(t) \xi_\beta(x) (\xi^\alpha)(-x)|\ 0>
+i<0\ |\theta(-t) \xi^\alpha(-x) \xi_\beta(x)|\ 0>\newline
=  +i<0\ |{\cal T} \xi^\alpha(-x) \xi_\beta(x)|\ 0>
=  <0\ |{\cal T} \xi^\alpha(-x) ((\eta_{\dot\beta})^c)^\dagger(x)|\ 0>.$

All these propagators are accordingly left invariant
\footnote{This is not much information, but it is correct.
Consider indeed the usual Feynman propagator in Fourier space for a Dirac
fermion with mass $m$
\begin{equation}
\int d^4x e^{ipx}<0\ |{\cal T} \left(\begin{array}{c} \xi^\alpha \cr
\eta_{\dot\alpha}\end{array}\right)(x) \left(\begin{array}{cc}
(\xi^\beta)^\dagger & (\eta_{\dot\beta})^\dagger\end{array}\right)(-x)\gamma^0
|\ 0> = \frac {p_\mu \gamma^\mu + m}{p^2 -m^2}
=\frac{1}{p^2-m^2}
\left(\begin{array}{cc} m & p_\mu\overline{\sigma^\mu} \cr
p_\mu\sigma^\mu & m \end{array}\right);
\label{eq:feyndir}
\end{equation}
it yields in particular (the $\gamma^0$ in (\ref{eq:feyndir})
 makes $\gamma^\mu_{\alpha,\beta}$ appear)
\begin{equation}
\int d^4x e^{ipx}<0\ |{\cal T} \xi^\alpha(x) \eta_{\dot\beta}(-x)|\ 0>
= \frac{p_\mu \gamma^\mu_{\alpha\beta}+ m\delta_{\alpha\beta}}{p^2 - m^2},
\alpha,\beta = 1,2.
\end{equation}
$PCT$ invariance tells us that, in a Dirac mass-like propagator,
 the $p^\mu$ term is not present, and the remaining term  is diagonal
in $\alpha,\beta$; and, indeed,
$\gamma^\mu_{\alpha\beta}$ vanishes $\forall \alpha,\beta = 1,2$, while the
term proportional to $m$ is diagonal in $\alpha,\beta$.}
 by the 4-inversion $x
\to -x$, or, in Fourier space, they are invariant when $p_\mu \to -p_\mu$.

\subsubsection{Constraints on kinetic-like terms}
%
\begin{eqnarray}
\ast\ Diagonal\
<0\ |{\cal T} \xi^\alpha(x) (\xi^\beta)^\dagger(-x)|\ 0>
&=&
-<0\ |{\cal T} \xi^\alpha(-x) (\xi^\beta)^\dagger(x)|\ 0>\cr
&=& <0\ |{\cal T} (\xi^\beta)^\dagger(x) \xi^\alpha(-x) |\ 0>;\cr
\ast\ Diagonal\
<0\ |{\cal T} (\xi^\alpha)^{c}(x) ((\xi^\beta)^c)^\dagger(-x)|\ 0>
&=&
-<0\ |{\cal T} (\xi^\alpha)^{c}(-x) ((\xi^\beta)^c)^\dagger(x)|\ 0>\cr
&=& <0\ |{\cal T} (\xi^\beta)^c)^\dagger(x)(\xi^\alpha)^{c}(-x) |\ 0>
;\cr
\ast\ Diagonal\
<0\ |{\cal T} (\eta_{\dot\alpha})^{c}(x) ((\eta_{\dot\beta})^c)^\dagger(-x)|\ 0>
&=&
-<0\ |{\cal T} (\eta_{\dot\alpha})^{c}(-x) ((\eta_{\dot\beta})^c)^\dagger(x)|\
0>\cr
&=& <0\ |{\cal T} ((\eta_{\dot\beta})^c)^\dagger(x)(\eta_{\dot\alpha})^{c}(-x) |\ 0>
;\cr
\ast\ Diagonal\
<0\ |{\cal T} \eta_{\dot\alpha}(x) (\eta_{\dot\beta})^\dagger(-x)|\ 0>
&=&
-<0\ |{\cal T} \eta_{\dot\alpha}(-x) (\eta_{\dot\beta})^\dagger(x)|\ 0>\cr
&=& <0\ |{\cal T} (\eta_{\dot\beta})^\dagger(x)  \eta_{\dot\alpha}(-x) |\ 0> ;\cr
&& \cr
\ast\ Non-diagonal\
<0\ |{\cal T} \xi^\alpha(x) ((\xi^\beta)^c)^\dagger(-x)|\ 0>
&=&
-<0\ |{\cal T} \xi^\alpha(-x) ((\xi^\beta)^c)^\dagger(x)|\ 0>\cr
&=& <0\ |{\cal T} ((\xi^\beta)^c)^\dagger(x) \xi^\alpha(-x)|\ 0>
;\cr
\ast\ Non-diagonal\
<0\ |{\cal T} (\xi^\alpha)^{c}(x) (\xi^\beta)^\dagger(-x)|\ 0>
&=&
-<0\ |{\cal T} (\xi^\alpha)^{c}(-x) (\xi^\beta)^\dagger(x)|\ 0>\cr
&=& <0\ |{\cal T} (\xi^\beta)^\dagger(x) (\xi^\alpha)^{c}(-x) |\ 0>
;\cr
\ast\ Non-diagonal\
<0\ |{\cal T} (\eta_{\dot\alpha})^{c}(x) (\eta_{\dot\beta})^\dagger(-x)|\ 0>
&=&
-<0\ |{\cal T} (\eta_{\dot\alpha})^{c}(-x) (\eta_{\dot\beta})^\dagger(x)|\ 0>\cr
&=& <0\ |{\cal T} (\eta_{\dot\beta})^\dagger(x) (\eta_{\dot\alpha})^{c}(-x) |\ 0>
;\cr
\ast\ Non-diagonal\
<0\ |{\cal T} \eta_{\dot\alpha}(x) ((\eta_{\dot\beta})^c)^\dagger(-x)|\ 0>
&=&
-<0\ |{\cal T} \eta_{\dot\alpha}(-x) ((\eta_{\dot\beta})^c)^\dagger(x)|\ 0>\cr
&=& <0\ |{\cal T} ((\eta_{\dot\beta})^c)^\dagger(x)\eta_{\dot\alpha}(-x) |\ 0>
.\cr
&&
\label{eq:cons2}
\end{eqnarray}
In Fourier space, all these propagators must accordingly be odd in $p_\mu$.
We check like above on the Dirac propagator that it is indeed the case.
One gets for example (the $\gamma^0$ in (\ref{eq:feyndir}) now makes
$\gamma^\mu_{\alpha,\beta+2}$ appear)
\begin{equation}
\int d^4x e^{ipx}<0\ |{\cal T} \xi^\alpha(x) (\xi^\beta)^\dagger(-x)|\ 0>
= \frac{p_\mu \gamma^\mu_{\alpha\beta+2}+ m\delta_{\alpha\beta+2}}{p^2 - m^2},
\alpha,\beta = 1,2,
\end{equation}
in which only the terms linear in $p_\mu$ are present, which are indeed odd
in $p_\mu$ as predicted by $PCT$ invariance.

Note that $PCT$ invariance does not forbid non-diagonal kinetic-like
propagators.

\subsubsection{Simple assumptions and consequences}

$PCT$ symmetry constrains, in Fourier space, all mass-like propagators to
be $p$-even and all kinetic-like propagators to be $p$-odd; the former
can only write $f(p^2)\delta_{\alpha\beta}$ and the latter
$g(p^2) p_\mu \sigma^\mu_{\alpha\beta}$ or $h(p^2) p_\mu
\overline{\sigma^\mu}_{\alpha\beta}$.

This is what we will suppose hereafter, and consider, in Fourier space,
a  propagator

\vbox{
\begin{eqnarray}
\Delta(p) &=&
\left(\begin{array}{cccc}
|\ \xi^\alpha> & |\ (\xi^\alpha)^c> & |\ (\eta_{\dot\alpha})^c> & |\
\eta_{\dot\alpha}>
\end{array}\right)\cr
&&\hskip -1cm\left(\begin{array}{ccc}
\left(\begin{array}{cc}\alpha_1(p^2) & a_1(p^2) \cr
b_1(p^2) & \beta_1(p^2) \end{array}\right)p_\mu \overline{\sigma^\mu}_{\alpha\beta}
 & \vline & \left( \begin{array}{cc} m_{L1}(p^2) & \mu_1(p^2) \cr
 m_1(p^2)  & m_{R1}(p^2)\end{array}\right)\delta_{\alpha\beta}  \cr
\hline
\left(\begin{array}{cc} m_{L2}(p^2) & m_2(p^2)\cr
\mu_2(p^2) & m_{R2}(p^2)\end{array}\right)\delta_{\alpha\beta} & \vline &
\left(\begin{array}{cc} \beta_2(p^2) & b_2(p^2) \cr
 a_2(p^2) & \alpha_2(p^2)  \end{array}\right)p_\mu {\sigma^\mu}_{\alpha\beta}
\end{array}\right)
\left(\begin{array}{c}
<\xi^\beta\ | \cr <(\xi^\beta)^c\ | \cr <(\eta_{\dot\beta})^c\ | \cr <
\eta_{\dot\beta}\ |
\end{array}\right).\cr
&&
\label{eq:propp}
\end{eqnarray}
}

This ans\"atz enables to get explicit constraints on the
propagator. It is motivated by the fact that, classically,
the (quadratic)
Lagrangian, which is the inverse propagator, has this same Lorentz
structure
\begin{equation}
L = \left(\begin{array}{ccc} K_1 (p_-)_{\alpha\beta} & \vline  & M_1\,
\delta_{\alpha\beta} \cr
\hline
M_2\, \delta_{\alpha\beta}  & \vline & K_2 (p_+)_{\alpha\beta}
\end{array}\right).
\end{equation}
An important property is that it automatically satisfies
the $PCT$ constraints (\ref{eq:cons1}) (\ref{eq:cons2}). For mass-like
propagators, which are invariant by the 4-inversion $x \to -x$ it is a
triviality; for kinetic like propagators, the ``$-$'' signs which occur in
the r.h.s.'s of (\ref{eq:cons2}) are canceled by the one which comes from
the differential operator $p_\mu$ acting on $(-x)$ instead of $x$.
We consider accordingly that
(\ref{eq:propp}) expresses the invariance of the propagator by $PCT$.

From now onwards we shall always use the form (\ref{eq:propp}) for the
propagator,  considering therefore that it is $PCT$ invariant.
It includes sixteen complex parameters.
We will see how individual discrete symmetries and their products
 reduce this number.

\subsection{Charge conjugate fields}
\label{subsection:CC}

By using the definitions of charge conjugate fields
\begin{eqnarray}
\xi^\alpha &=& g^{\alpha\gamma}\xi_\gamma = -i\sigma^2_{\alpha\gamma}
\xi_\gamma = -i\sigma^2_{\alpha\gamma}(-i)((\eta_{\dot\gamma})^c)^\dagger
=-\sigma^2_{\alpha\gamma}((\eta_{\dot\gamma})^c)^\dagger,\cr
\eta_{\dot\beta} &=& g_{\beta\delta}\eta^{\dot\delta}
=i\sigma^2_{\beta\delta}\eta^{\dot\delta}
=i\sigma^2_{\beta\delta}(-i)((\xi^\delta)^c)^\dagger
=\sigma^2_{\beta\delta}((\xi^\delta)^c)^\dagger.
\label{eq:Cdefi}
\end{eqnarray}
one can bring additional constraints to the ones obtained from expressing
the invariance by a discrete symmetry like $PCT$.
We first give the example of a Dirac-like propagator:

$<0\ |{\cal T} \xi^\alpha(x) (\eta_{\dot\beta})^\dagger(-x)|\ 0>
= <0\ |{\cal T} (-)\sigma^2_{\alpha\gamma}((\eta_{\dot\gamma})^c)^\dagger(x)
\left(\sigma^2_{\beta\delta}((\xi^\delta)^c)^\dagger(-x)\right)^\dagger|\
0>\\
= \sigma^2_{\alpha\gamma}\sigma^2_{\beta\delta}
<0\ | {\cal T}((\eta_{\dot\gamma})^c)^\dagger(x)(\xi^\delta)^c(-x)|\ 0>
= (\delta_{\alpha\delta} \delta_{\beta\gamma} - \delta_{\alpha\beta}
\delta_{\delta\gamma})
<0\ | {\cal T}((\eta_{\dot\gamma})^c)^\dagger(x)(\xi^\delta)^c(-x)|\ 0>
= - <0\ | {\cal T} (\xi^\alpha)^c(-x) ((\eta_{\dot\beta})^c)^\dagger(x)|\ 0>
+\delta_{\alpha\beta}
<0\ | {\cal T} (\xi^\gamma)^c(-x) ((\eta_{\dot\gamma})^c)^\dagger(x)|\ 0>$.

The r.h.s. of the corresponding $PCT$ constraint in the first line of
(\ref{eq:cons1}) writes the same but for the exchange $x \to (-x)$.
If we now use the ans\"atz (\ref{eq:propp}) which implements $PCT$
invariance, one gets
\begin{equation}
\mu_1(p^2)\delta_{\alpha\beta} =
-(\delta_{\beta\gamma\delta_\alpha\delta} -
\delta_{\alpha\beta}\delta_{\delta\gamma})
m_1(p^2) \delta_{\delta\gamma}
= \delta_{\alpha\beta} m_1(p^2),
\end{equation}
equivalently
\begin{equation}
m_1(p^2) = \mu_1(p^2).
\end{equation}
Likewise, one gets $m_2(p^2) = \mu_2(p^2)$.

For Majorana-like propagator, using the definitions (\ref{eq:Cdefi})
of charge conjugate fields, one gets
\begin{eqnarray}
<0\ |\ {\cal T} \xi^\alpha(x) (\eta_{\dot\beta}^c)^\dagger(-x)\ |\ 0>
&=& <0 \ |\ {\cal T}(\eta_{\dot\beta}^c)^\dagger(x) \xi^\alpha(-x)\ |\ 0>
-\delta_{\alpha\beta}
<0\ |\ {\cal T} (\eta_{\dot\gamma}^c)^\dagger(x) \xi^\gamma(-x)\ |\ 0>\cr
&=& - <0\ |\ {\cal T} \xi^\alpha(-x) (\eta_{\dot\beta}^c)^\dagger(x)\ |\ 0>
+\delta_{\alpha\beta}
 <0\ |\ {\cal T} \xi^\gamma(-x) (\eta_{\dot\gamma}^c)^\dagger(x)\ |\ 0>,\cr
&&
\label{eq:majcon1}
\end{eqnarray}
while, with the same procedure, its transformed by $PCT$ in the r.h.s. of
(\ref{eq:cons1}) becomes
\begin{eqnarray}
-<0\ |\ {\cal T} (\eta_{\dot\beta}^c)^\dagger(x) \xi^\alpha(-x)\ |\ 0>
&=&
-<0\ |\ {\cal T} \xi^\alpha(x) (\eta{\dot\beta}^c)^\dagger(-x) +
\delta_{\alpha\beta}
<0\ |\ {\cal T} \xi^\gamma(x) (\eta_{\dot\gamma}^c)^\dagger(-x)\ |\ 0>.\cr
&&
\label{eq:majcon2}
\end{eqnarray}
One only gets tautologies such that no additional constraint arises.

We implement the same procedure for kinetic-like terms, for example
$<0\ |{\cal T} \xi^\alpha(x) (\xi^\beta)^\dagger(-x)|\ 0> =
<0\ |{\cal T} (\xi^\beta)^\dagger (x) \xi^\alpha(-x)|\ 0>$.
Using $\xi^\alpha =
-\sigma^2_{\alpha\gamma}((\eta_{\dot\gamma})^c)^\dagger$ and
$(\xi^\beta)^\dagger = \sigma^2_{\beta\delta}(\eta_{\dot\delta})^c$ and
(\ref{eq:propp}), one gets
\begin{eqnarray}
\alpha_1(p^2) p_\mu\overline{\sigma^\mu}_{\alpha\beta}
&=& -(\delta_{\beta\gamma\delta_\alpha\delta} -
\delta_{\alpha\beta}\delta_{\delta\gamma})
\beta_2(p^2) p_\mu{\sigma^\mu}_{\delta\gamma}\cr
&=& -\beta_2(p^2)(p_\mu{\sigma^\mu}_{\alpha\beta}
-\delta_{\alpha\beta} p_\mu Tr{\sigma_\mu})\cr
&=& -\beta_2(p^2)\left(p_\mu{\sigma^\mu}_{\alpha\beta}
-\delta_{\alpha\beta} (2 p_0 + 0\times p^i)\right)\cr
&=&
-\beta_2(p^2)(-p_0\sigma^0_{\alpha\beta}+\vec p.\vec\sigma_{\alpha\beta})\cr
&=&\beta_2(p^2)p_\mu\overline{\sigma^\mu}_{\alpha\beta},
\end{eqnarray}
which entails
\begin{equation}
\alpha_1(p^2) = \beta_2(p^2).
\end{equation}
Likewise, one gets $\alpha_2(p^2) = \beta_1(p^2)$, and, for the
non-diagonal kinetic-like propagators, $a_1(p^2)=a_2(p^2), b_1(p^2)
=b_2(p^2)$.

So, after making use of the definition of charge conjugate fields,
 (\ref{eq:propp}) expressing the $PCT$ invariance of the propagator
 rewrites

\vbox{
\begin{eqnarray}
\Delta_{PCT}(p) &=&
\left(\begin{array}{cccc}
|\ \xi^\alpha> & |\ (\xi^\alpha)^c> & |\ (\eta_{\dot\alpha})^c> & |\
\eta_{\dot\alpha}>
\end{array}\right)\cr
&&\hskip -1cm\left(\begin{array}{ccc}
\left(\begin{array}{cc}\alpha(p^2) & u(p^2) \cr
v(p^2) & \beta(p^2) \end{array}\right)p_\mu \overline{\sigma^\mu}_{\alpha\beta}
 & \vline & \left( \begin{array}{cc} m_{L1}(p^2) & \mu_1(p^2) \cr
 \mu_1(p^2)  & m_{R1}(p^2)\end{array}\right)\delta_{\alpha\beta}  \cr
\hline
\left(\begin{array}{cc} m_{L2}(p^2) & \mu_2(p^2)\cr
\mu_2(p^2) & m_{R2}(p^2)\end{array}\right)\delta_{\alpha\beta} & \vline &
\left(\begin{array}{cc} \alpha(p^2) & v(p^2) \cr
 u(p^2) & \beta(p^2)  \end{array}\right)p_\mu {\sigma^\mu}_{\alpha\beta}
\end{array}\right)
\left(\begin{array}{c}
<\xi^\beta\ | \cr <(\xi^\beta)^c\ | \cr <(\eta_{\dot\beta})^c\ | \cr <
\eta_{\dot\beta}\ |
\end{array}\right).\cr
&&
\label{eq:propPCT}
\end{eqnarray}
}

$PCT$ symmetry has finally reduced the total number of arbitrary functions
 necessary to describe one flavor of fermions from 16 to 10.

\subsection{$\boldsymbol{C}$ constraints}
\label{subsection:Cprop}

$C$ is a unitary operator and we may use directly (\ref{eq:psiopC}) in the
expression of the propagator.
This is an example of demonstration, in which we suppose that the vacuum is
invariant by $C$.

$<0\ |{\cal T} \xi^\alpha(x) (\eta_{\dot\beta})^\dagger(-x)|\ 0>
= <C\, 0\ | {\cal T} C(-i(\eta^{\dot\alpha})^\dagger)(x) C^{-1}
C(i\xi_\beta)(-x) C^{-1}|\ C\, 0> \newline
= <C\, 0\ |{\cal T} C (\eta^{\dot\alpha})^\dagger)(x) \xi_\beta(-x) C^{-1}|
\ C\, 0> = <0\ | {\cal T} C^\dagger C (\eta^{\dot\alpha})^\dagger)(x)
\xi_\beta(-x) C^{-1} C\ |\ 0> \newline
= <0\ |{\cal T} (\eta^{\dot\alpha})^\dagger)(x) \xi_\beta(-x)|\ 0>
= <0\ |{\cal T} ((\xi^\alpha)^c)^\dagger (x)
((\eta_{\dot\beta})^c)^\dagger(-x)|\ 0>. $

By using (\ref{eq:propp}) expressing $PCT$ invariance, one gets accordingly

\vbox{
\begin{eqnarray}
\Delta_{C + PCT}(p) &=&
\left(\begin{array}{cccc}
|\ \xi^\alpha> & |\ (\xi^\alpha)^c> & |\ (\eta_{\dot\alpha})^c> & |\
\eta_{\dot\alpha}>
\end{array}\right)\cr
&&\hskip -1cm\left(\begin{array}{ccc}
\left(\begin{array}{cc}\alpha(p^2) & a(p^2)\cr
a(p^2) & \alpha(p^2)\end{array}\right) p_\mu\overline{\sigma^\mu}_{\alpha\beta} & \vline &
\left(\begin{array}{cc} \rho(p^2) & \mu(p^2) \cr
\mu(p^2)  & \rho(p^2)\end{array}\right)\delta_{\alpha\beta} \cr
\hline
\left(\begin{array}{cc}\sigma(p^2) & m(p^2) \cr
m(p^2) & \sigma(p^2)\end{array}\right) \delta_{\alpha\beta}
 & \vline & \left(\begin{array}{cc} \beta(p^2) & b(p^2) \cr
 b(p^2) & \beta(p^2)\end{array}\right) p_\mu{\sigma^\mu}_{\alpha\beta}
\end{array}\right)
\left(\begin{array}{c}
<\xi^\beta\ | \cr <(\xi^\beta)^c\ | \cr <(\eta_{\dot\beta})^c\ | \cr <
\eta_{\dot\beta}\ |
\end{array}\right).\cr
&&
\label{eq:Cprop1}
\end{eqnarray}
}

All $2\times 2$ submatrices are in particular symmetric.

Combining now (\ref{eq:propPCT}) and (\ref{eq:Cprop1}), a
 $C$ + $PCT$ invariant propagator, after using the definition of charge
conjugate fields, can finally be reduced to

\vbox{
\begin{eqnarray}
\Delta_{C+PCT}(p) &=&
\left(\begin{array}{cccc}
|\ \xi^\alpha> & |\ (\xi^\alpha)^c> & |\ (\eta_{\dot\alpha})^c> & |\
\eta_{\dot\alpha}>
\end{array}\right)\cr
&&\hskip -1cm\left(\begin{array}{ccc}
\left(\begin{array}{cc}\alpha(p^2) & a(p^2)\cr
a(p^2)  & \alpha(p^2)\end{array}\right)p_\mu\overline{\sigma^\mu}_{\alpha\beta}
 & \vline &
\left(\begin{array}{cc}\rho(p^2) & \mu(p^2) \cr
\mu(p^2)  & \rho(p^2)\end{array}\right)\delta_{\alpha\beta} \cr
\hline
\left(\begin{array}{cc}\sigma(p^2) & m(p^2)\cr
m(p^2) & \sigma(p^2)\end{array}\right)\delta_{\alpha\beta}
& \vline &
\left(\begin{array}{cc}\alpha(p^2) & a(p^2) \cr
a(p^2)  & \alpha(p^2)\end{array}\right)p_\mu{\sigma^\mu}_{\alpha\beta}
\end{array}\right)
\left(\begin{array}{c}
<\xi^\beta\ | \cr <(\xi^\beta)^c\ | \cr <(\eta_{\dot\beta})^c\ | \cr <
\eta_{\dot\beta}\ |
\end{array}\right),\cr
&&
\label{eq:CPCTprop1}
\end{eqnarray}
}

in which the number of arbitrary functions has now been reduced to 6.

\subsection{$\boldsymbol{P}$ constraints}
\label{subsection:Pprop}

In momentum space, the parity transformed of $p_\mu\sigma^\mu \equiv
(p_0\sigma^0 + \vec p. \vec\sigma)$ is $(p_0\sigma^0 -\vec p.\vec\sigma)
 \equiv p_\mu\overline{\sigma^\mu}$.

Using (\ref{eq:Plandop}) and the assumption (\ref{eq:propp}) expressing
$PCT$ invariance,
and supposing the vacuum invariant by parity, one gets

\vbox{
\begin{eqnarray}
\Delta_{P + PCT}(p) &=&
\left(\begin{array}{cccc}
|\ \xi^\alpha> & |\ (\xi^\alpha)^c> & |\ (\eta_{\dot\alpha})^c> & |\
\eta_{\dot\alpha}>
\end{array}\right)\cr
&&\left(\begin{array}{ccc}
\left(\begin{array}{cc}\alpha(p^2) & a(p^2)\cr
b(p^2) & \beta(p^2)\end{array}\right)p_\mu\overline{\sigma^\mu}_{\alpha\beta}
& \vline &
\left(\begin{array}{cc}\rho(p^2) & \mu(p^2) \cr
 m(p^2)  & \sigma(p^2)\end{array}\right)\delta_{\alpha\beta} \cr
\hline
\left(\begin{array}{cc}\sigma(p^2) & m(p^2)\cr
\mu(p^2) & \rho(p^2)\end{array}\right)\delta_{\alpha\beta}
& \vline &
\left(\begin{array}{cc}\beta(p^2) & b(p^2) \cr
 a(p^2) & \alpha(p^2)\end{array}\right)p_\mu{\sigma^\mu}_{\alpha\beta}
\end{array}\right)
\left(\begin{array}{c}
<\xi^\beta\ | \cr <(\xi^\beta)^c\ | \cr <(\eta_{\dot\beta})^c\ | \cr <
\eta_{\dot\beta}\ |
\end{array}\right).\cr
&&
\label{eq:Pprop0}
\end{eqnarray}
}

A $P$ + $C$ + $PCT$ invariant propagator writes

\vbox{
\begin{eqnarray}
\Delta_{P+C + PCT}(p) &=&
\left(\begin{array}{cccc}
|\ \xi^\alpha> & |\ (\xi^\alpha)^c> & |\ (\eta_{\dot\alpha})^c> & |\
\eta_{\dot\alpha}>
\end{array}\right)\cr
&& \hskip -5mm\left(\begin{array}{ccccc}
\left(\begin{array}{cc}\alpha(p^2) & a(p^2)\cr
a(p^2) & \alpha(p^2)\end{array}\right)p_\mu\overline{\sigma^\mu}_{\alpha\beta}
& \vline &
\left(\begin{array}{cc}\rho(p^2) & \mu(p^2) \cr
\mu(p^2)  & \rho(p^2)\end{array}\right)\delta_{\alpha\beta} \cr
\hline
\left(\begin{array}{cc}\rho(p^2) & \mu(p^2)\cr
\mu(p^2) & \rho(p^2)\end{array}\right)\delta_{\alpha\beta}
& \vline &
\left(\begin{array}{cc}\alpha(p^2) & a(p^2)\cr
a(p^2) & \alpha(p^2)\end{array}\right)
p_\mu{\sigma^\mu}_{\alpha\beta}
\end{array}\right)
\left(\begin{array}{c}
<\xi^\beta\ | \cr <(\xi^\beta)^c\ | \cr <(\eta_{\dot\beta})^c\ | \cr <
\eta_{\dot\beta}\ |
\end{array}\right).\cr
&&
\label{eq:P+Cprop1}
\end{eqnarray}
}

The expressions above can be further reduced by using the definition of
charge conjugate fields, which leads to (\ref{eq:propPCT}) as the
expression of $PCT$ invariance. So doing,
a $P$ + $PCT$ invariant propagator writes

\vbox{
\begin{eqnarray}
\Delta_{P+PCT}(p) &=&
\left(\begin{array}{cccc}
|\ \xi^\alpha> & |\ (\xi^\alpha)^c> & |\ (\eta_{\dot\alpha})^c> & |\
\eta_{\dot\alpha}>
\end{array}\right)\cr
&&\hskip -1cm\left(\begin{array}{ccccc}
\left(\begin{array}{cc}\alpha(p^2) & a(p^2)\cr
b(p^2) & \alpha(p^2)\end{array}\right)p_\mu\sigma^\mu_{\alpha\beta}
 & \vline &
\left(\begin{array}{cc}\rho(p^2) & \mu(p^2)  \cr
 \mu(p^2)  & \sigma(p^2)\end{array}\right)\delta_{\alpha\beta} \cr
\hline
\left(\begin{array}{cc}\sigma(p^2) & \mu(p^2)\cr
\mu(p^2) & \rho(p^2)\end{array}\right)\delta_{\alpha\beta}
 & \vline &
\left(\begin{array}{cc}\alpha(p^2) & b(p^2) \cr
a(p^2)  & \alpha(p^2)\end{array}\right)p_\mu\overline{\sigma^\mu}_{\alpha\beta}
\end{array}\right)
\left(\begin{array}{c}
<\xi^\beta\ | \cr <(\xi^\beta)^c\ | \cr <(\eta_{\dot\beta})^c\ | \cr <
\eta_{\dot\beta}\ |
\end{array}\right);\cr
&&
\label{eq:PPCTprop1}
\end{eqnarray}
}

and one finds again the expression (\ref{eq:P+Cprop1}) for a $P + C + PCT$
invariant propagator.

\subsection{$\boldsymbol{CP}$ constraints}
\label{subsection:CPprop}

Using (\ref{eq:CPlag1}), (\ref{eq:propp}),
 and supposing the vacuum invariant by $CP$, one gets

\vbox{
\begin{eqnarray}
\Delta_{CP + PCT}(p) &=&
\left(\begin{array}{cccc}
|\ \xi^\alpha> & |\ (\xi^\alpha)^c> & |\ (\eta_{\dot\alpha})^c> & |\
\eta_{\dot\alpha}>
\end{array}\right)\cr
&&\left(\begin{array}{ccc}
\left(\begin{array}{cc}\alpha(p^2) & u(p^2)\cr
v(p^2) & \beta(p^2)\end{array}\right)p_\mu\overline{\sigma^\mu}_{\alpha\beta}
& \vline &
\left(\begin{array}{cc}m_L(p^2) & \mu(p^2) \cr
m(p^2)  & m_R(p^2)\end{array}\right)\delta_{\alpha\beta} \cr
\hline
\left(\begin{array}{cc}m_L(p^2) & \mu(p^2)\cr
m(p^2) & m_R(p^2)\end{array}\right)\delta_{\alpha\beta}
& \vline &
\left(\begin{array}{cc}\alpha(p^2) & u(p^2) \cr
v(p^2) & \beta(p^2)\end{array}\right) p_\mu{\sigma^\mu}_{\alpha\beta}
\end{array}\right)
\left(\begin{array}{c}
<\xi^\beta\ | \cr <(\xi^\beta)^c\ | \cr <(\eta_{\dot\beta})^c\ | \cr <
\eta_{\dot\beta}\ |
\end{array}\right).\cr
&&
\label{eq:CPprop1}
\end{eqnarray}
}

It can be further constrained by using the definition of charge conjugate
fields which makes the $PCT$ constraint be (\ref{eq:propPCT}), to

\vbox{
\begin{eqnarray}
\Delta_{CP+PCT}(p) &=&
\left(\begin{array}{cccc}
|\ \xi^\alpha> & |\ (\xi^\alpha)^c> & |\ (\eta_{\dot\alpha})^c> &
|\ \eta_{\dot\alpha}>
\end{array}\right)\cr
&&\hskip -1cm\left(\begin{array}{ccccc}
\left(\begin{array}{cc}\alpha(p^2) & u(p^2)\cr
u(p^2) & \beta(p^2)\end{array}\right)p_\mu\overline{\sigma^\mu}_{\alpha\beta}
& \vline &
\left(\begin{array}{cc} m_L(p^2) & \mu(p^2) \cr
\mu(p^2)  & m_R(p^2) \end{array}\right)\delta_{\alpha\beta} \cr
\hline
\left(\begin{array}{cc} m_L(p^2)  &  \mu(p^2)\cr
\mu(p^2) & m_R(p^2) \end{array}\right)\delta_{\alpha\beta}
& \vline &
\left(\begin{array}{cc}\alpha(p^2) & u(p^2) \cr
u(p^2) & \beta(p^2)\end{array}\right)
p_\mu{\sigma^\mu}_{\alpha\beta}
\end{array}\right)
\left(\begin{array}{c}
<\xi^\beta\ | \cr <(\xi^\beta)^c\ | \cr <(\eta_{\dot\beta})^c\ | \cr
<\eta_{\dot\beta}\ |
\end{array}\right).\cr
&&
\label{eq:CPPCTprop2}
\end{eqnarray}
}

One then gets 4 symmetric $2 \times 2$ sub-blocks.

\subsection{Eigenstates of a  $\boldsymbol{C+PCT}$ invariant propagator}
\label{subsection:CPCTeig}

We do not consider any $PCT$ violation, because, if
this occurred, the very foundations of local Quantum Field Theory would be
undermined, and the meaning of our conclusions itself could thus strongly
be cast in doubt.

We look here for the eigenstates of the $4\times 4$ matrix in
 (\ref{eq:CPCTprop1})
\begin{equation}
\Delta_{C+PCT}(p^2) =
\left(\begin{array}{ccccc}
\left(\begin{array}{cc}\alpha(p^2) & a(p^2)\cr
  a(p^2)& \alpha(p^2)\end{array}\right)p_\mu\sigma^\mu_{\alpha\beta}
 & \vline &
\left(\begin{array}{cc}\rho(p^2) & \mu(p^2) \cr
\mu(p^2)  & \rho(p^2)\end{array}\right)\delta_{\alpha\beta} \cr
\hline
\left(\begin{array}{cc}\sigma(p^2) & m(p^2)\cr
m(p^2) & \sigma(p^2)\end{array}\right)\delta_{\alpha\beta}
& \vline &
\left(\begin{array}{cc}\alpha(p^2) & a(p^2) \cr
 a(p^2) & \alpha(p^2)\end{array}\right)p_\mu\overline{\sigma^\mu}_{\alpha\beta}
\end{array}\right).
\label{eq:CPCTprop2}
\end{equation}

The three symmetric matrices $\left(\begin{array}{cc} \rho & \mu \cr \mu & \rho
\end{array}\right)$, $\left(\begin{array}{rr} \sigma & m \cr m &
\sigma \end{array}\right)$ and $\left(\begin{array}{cc} \alpha & a \cr
a & \alpha \end{array}\right)$
 can be simultaneously diagonalized by a unitary matrix
$U$ according to

\vbox{
\begin{eqnarray}
U^T \left(\begin{array}{cc} \rho & \mu \cr \mu & \rho
\end{array}\right) U &=& \left(\begin{array}{cc} (\rho+\mu)e^{2i\varphi} &\cr
& (\rho-\mu)e^{-2i\varphi}\end{array}\right),\cr
U^T \left(\begin{array}{rr} \alpha & a \cr a & \alpha
\end{array}\right) U &=& \left(\begin{array}{cc} (\alpha+a)e^{2i\varphi} &\cr
& (\alpha-a)e^{-2i\varphi}\end{array}\right),\cr
U&=&\frac{1}{\sqrt{2}}e^{i\omega}\left(\begin{array}{rr}
e^{i\varphi} & -e^{-i\varphi} \cr e^{i\varphi} &
e^{-i\varphi}\end{array}\right).
\end{eqnarray}
}

We can choose the particular case
\begin{equation}
U=U_0\equiv\frac{1}{\sqrt{2}}\left(\begin{array}{rr}
1 & -1 \cr 1 & 1\end{array}\right).
\label{eq:U0}
\end{equation}

Call the initial basis
\begin{equation}
<n_L\ | = \left(\begin{array}{c} <\xi^\alpha\ | \cr <(\xi^\beta)^c\ |
\end{array}\right)
\equiv
\left(\begin{array}{c} <\xi^\alpha\ | \cr <-i(\eta^{\dot\beta})^\dagger\ |
\end{array}\right)
,\quad
<n_R\ | = \left(\begin{array}{c} <(\eta_{\dot\alpha})^c\ | \cr
<\eta_{\dot\beta}\ |
\end{array}\right)
\equiv \left(\begin{array}{c} <-i(\xi_\alpha)^\dagger\ | \cr
<\eta_{\dot\beta}\ |
\end{array}\right),
\end{equation}
one has
\begin{equation}
\left(\begin{array}{cccc}
|\ \xi^\alpha> & |\ (\xi^\beta)^c> & |\ (\eta_{\dot\gamma})^c> & |\
\eta_{\dot\delta}>
\end{array}\right)
=\left(\begin{array}{cc} |\ n_L> & |\ n_R>\end{array}\right).
\end{equation}
Define the new basis by
\begin{eqnarray}
<N_L\ | = U_0^\dagger <n_L\ |&,& <N_R\ |= U_0^\dagger <n_R\ |,\cr
|\ N_L>= U_0\;|\ n_L> &,& |\ N_R> = U_0\; |\ n_R>.
\end{eqnarray}
One has explicitly
\begin{eqnarray}
<N_L\ |&=&\frac{1}{\sqrt{2}}\left(\begin{array}{c}
<\xi^\alpha -i (\eta^{\dot\alpha})^\dagger\ |\cr
<-\xi^\alpha-i (\eta^{\dot\alpha})^\dagger\ |\end{array} \right)
=\frac{1}{\sqrt{2}}\left(\begin{array}{c}
<\xi^\alpha +(\xi^\alpha)^c\ |\cr
<-\xi^\alpha + (\xi^\alpha)^c\ |\end{array} \right),\cr
<N_R\ |&=&\frac{1}{\sqrt{2}}\left(\begin{array}{c}
<-i(\xi_\alpha)^\dagger + \eta_{\dot\alpha}\ |\cr
<+i(\xi_\alpha)^\dagger + \eta_{\dot\alpha}\ |\end{array}\right)
=\frac{1}{\sqrt{2}}\left(\begin{array}{c}
<\eta_{\dot\alpha}+ (\eta_{\dot\alpha})^c\ |\cr
<\eta_{\dot\alpha}- (\eta_{\dot\alpha})^c\ |\end{array}\right),
\end{eqnarray}
and one can write
\begin{equation}
<N_L\ |= \left(\begin{array}{c} <\chi^\alpha\ | \cr
<(-i)(\omega^{\dot\beta})^\dagger\ | \end{array}\right),
<N_R\ | = \left(\begin{array}{c} <(-i) (\chi_\alpha)^\dagger\ | \cr
<\omega_{\dot\beta}\ | \end{array}\right).
\end{equation}
In this new basis, the propagator writes (using
(from (\ref{eq:U0})) $U_0^T U_0 = 1$)

\vbox{
\begin{eqnarray}
&&\Delta_{C+PCT}(p^2)=\left(\begin{array}{cc} |\ N_L> & |\ N_R>
\end{array}\right)\cr
&&
\hskip -2cm
\left(\begin{array}{ccccc}
\left(\begin{array}{cc}\alpha(p^2)+a(p^2) & \cr
  & \alpha(p^2)-a(p^2)\end{array}\right)p_\mu\sigma^\mu_{\alpha\beta}
& \vline &
\left(\begin{array}{cc}\rho(p^2)+\mu(p^2) &  \cr
& \rho(p^2)-\mu(p^2)\end{array}\right)\delta_{\alpha\beta} \cr
\hline
\left(\begin{array}{cc}\sigma(p^2)+m(p^2) & \cr
& \sigma(p^2)-m(p^2)\end{array}\right)\delta_{\alpha\beta}
& \vline &
\left(\begin{array}{cc}\alpha(p^2)+a(p^2) &  \cr
& \alpha(p^2)-a(p^2)\end{array}\right)p_\mu\overline{\sigma^\mu}_{\alpha\beta}
\end{array}\right)
\left(\begin{array}{c}  <N_L\ | \cr  <N_R\ |
\end{array}\right).\cr
&&
\label{eq:Cmaj0}
\end{eqnarray}
}
Remember that $|\ u><v\ |$ corresponds, in our notation, to a propagator
$<0\ |{\cal T} u(x) v^\dagger(-x)|\ 0>$.

One introduces the  Majorana fermions (see subsection
\ref{subsection:majorana})
\begin{eqnarray}
X^\pm_M &=& \left(\begin{array}{c} \chi^\alpha \cr
\pm(-i)(\chi_\alpha)^\dagger \end{array}\right)
=\frac{1}{\sqrt{2}}\left(\begin{array}{c}
\xi^\alpha +(\xi^\alpha)^c \cr
\pm \left(\eta_{\dot\alpha} + (\eta_{\dot\alpha})^c\right)
\end{array}\right)
=\frac{1}{\sqrt{2}}\left(\begin{array}{c}
\xi^\alpha -i(\eta^{\dot\alpha})^\dagger \cr
\pm (\eta_{\dot\alpha} -i(\xi_\alpha)^\dagger)
\end{array}\right),
\cr
\Omega^\pm_M &=& \left(\begin{array}{c} \pm(-i)(\omega^{\dot\beta})^\dagger \cr
\omega_{\dot\beta} \end{array}\right)
=\frac{1}{\sqrt{2}}\left(\begin{array}{c} \pm\left(-\xi^\beta
+(\xi^\beta)^c \right) \cr
\eta_{\dot\beta}-(\eta_{\dot\beta})^c \end{array}\right)
=\frac{1}{\sqrt{2}}\left(\begin{array}{c} \pm(-\xi^\beta
-i(\eta^{\dot\beta})^\dagger ) \cr
\eta_{\dot\beta}+i(\xi_\beta)^\dagger \end{array}\right).\cr
&&
\label{eq:majo1}
\end{eqnarray}

\subsubsection{Kinetic-like propagators}
%
They can be rewritten
\begin{eqnarray}
\int d^4x e^{ipx} <0\ |{\cal T} \chi^\alpha (x) (\chi^\beta)^\dagger  (-x)|\ 0>
&=& (\alpha(p^2) +a(p^2))p_\mu \sigma^\mu_{\alpha\beta},\cr
\int d^4x e^{ipx}<0\ |{\cal T} (\chi_\alpha)^\dagger (x) \chi_\beta (-x)|\ 0> &=&
(\alpha(p^2) + a(p^2))p_\mu \overline{\sigma^\mu}_{\alpha\beta},\cr
\int d^4x e^{ipx} <0\ |{\cal T}  (\omega^{\dot\alpha})^\dagger(x)
\omega^{\dot\beta}   (-x)|\ 0>
&=&   (\alpha(p^2)-a(p^2))p_\mu \sigma^\mu_{\alpha\beta} ,\cr
\int d^4x e^{ipx} <0\ |{\cal T} \omega_{\dot\alpha}
(x)(\omega_{\dot\beta})^\dagger   (-x)|\ 0> &=&
(\alpha(p^2)-a(p^2))p_\mu \overline{\sigma^\mu}_{\alpha\beta},
\end{eqnarray}

\subsubsection{Mass-like propagators}

They write
\begin{eqnarray}
\int d^4x e^{ipx} <0\ |{\cal T} \chi^\alpha(x) i\chi_\beta(-x)|\ 0> &=&
\delta_{\alpha\beta} (\rho(p^2)+\mu(p^2)),\cr
\int d^4x e^{ipx} <0\ |{\cal T} (-i)(\chi_\alpha)^\dagger(x)
(\chi^\beta)^\dagger (-x)|\ 0> &=&
\delta_{\alpha\beta} (\sigma(p^2)+m(p^2)),\cr
\int d^4x e^{ipx} <0\ |{\cal T}
(-i)(\omega^{\dot\alpha})^\dagger(x)(\omega_{\dot\beta})^\dagger(-x)|\
0> &=& \delta_{\alpha\beta} (\rho(p^2)-\mu(p^2)),\cr
\int d^4x e^{ipx} <0\ |{\cal T}\omega_{\dot\alpha} (x)  i\omega^{\dot\beta} (-x)|\ 0>
&=&
\delta_{\alpha\beta} (\sigma(p^2)-m(p^2)).
\end{eqnarray}

\subsubsection{Conclusion}

When $C$ and $PCT$ invariance holds, the fermion propagator decomposes into
the propagators for the Majorana fermions $X$ and $\Omega$ (\ref{eq:majo1})
(note that we have introduced below the $\overline{( )}$ fields instead
of the $( )^\dagger$ fields, thus an extra $\gamma^0$ matrix)

\vbox{
\begin{eqnarray}
&&\int d^4x e^{ipx}<0\ |{\cal T} {X^\pm_{M\alpha}}(x) \overline{X^\pm_{M\beta}}(-x)|\ 0>
=\left(\begin{array}{cc} (\rho(p^2)+\mu(p^2))\delta_{\alpha\beta} &
(\alpha(p^2)+a(p^2))p_\mu\sigma^\mu_{\alpha\beta} \cr
(\alpha(p^2)+a(p^2))p_\mu\overline{\sigma^\mu}_{\alpha\beta} &(\sigma(p^2) +
m(p^2))\delta_{\alpha\beta}
\end{array}\right),\cr
&&\int d^4x e^{ipx}<0\ |{\cal T} {\Omega^\pm_{M\alpha}}(x)
\overline{\Omega^\pm_{M\beta}}(-x)|\ 0>=
\left(\begin{array}{cc}(\rho(p^2) - \mu(p^2))\delta_{\alpha\beta} &
 (\alpha(p^2)-a(p^2))p_\mu\sigma^\mu_{\alpha\beta}  \cr
(\alpha(p^2)-a(p^2))p_\mu\overline{\sigma^\mu}_{\alpha\beta} &
(\sigma(p^2) - m(p^2))\delta_{\alpha\beta}
\end{array}\right).\cr
&&
\label{eq:Cmaj1}
\end{eqnarray}
}

(\ref{eq:Cmaj1}) also writes

\vbox{
\begin{eqnarray}
&&\frac{1}{2}\int d^4x e^{ipx}\left(<0\ |{\cal T} {X^\pm_{M\alpha}}(x)
\overline{X^\pm_{M\beta}}(-x)|\ 0> +<0\ |{\cal T} {\Omega^\pm_{M\alpha}}(x)
\overline{\Omega^\pm_{M\beta}}(-x)|\ 0>
\right)\cr
&&\hskip 6cm =
\left(\begin{array}{cc} \rho(p^2)\delta_{\alpha\beta} &
\alpha(p^2)p_\mu\sigma^\mu_{\alpha\beta} \cr
\alpha(p^2)p_\mu\overline{\sigma^\mu}_{\alpha\beta} &\sigma(p^2)
\delta_{\alpha\beta}
\end{array}\right),\cr
&&\frac{1}{2}\int d^4x e^{ipx}\left(<0\ |{\cal T} {X^\pm_{M\alpha}}(x)
\overline{X^\pm_{M\beta}}(-x)|\ 0> - <0\ |{\cal T} {\Omega^\pm_{M\alpha}}(x)
\overline{\Omega^\pm_{M\beta}}(-x)|\ 0>\right)\cr
&&\hskip 6cm =
\left(\begin{array}{cc}\mu(p^2)\delta_{\alpha\beta} &
 a(p^2)p_\mu\sigma^\mu_{\alpha\beta}  \cr
a(p^2)p_\mu\overline{\sigma^\mu}_{\alpha\beta} &
 m(p^2)\delta_{\alpha\beta}
\end{array}\right).\cr
&&
\label{eq:Cmaj2}
\end{eqnarray}
}

So, when $C$ + $PCT$ invariance is realized,  the most general fermion
propagator is equivalent to two Majorana propagators.

The determinant of $\Delta(p^2)$ (\ref{eq:Cmaj0}) is the products of the
determinants of the matrices in the r.h.s. of (\ref{eq:Cmaj1}); so, the
poles of the two Majorana propagators in (\ref{eq:Cmaj1}) are also poles of
$\Delta(p^2)$, and the physical states (eigenstates of the propagator at
its poles) are the Majorana fermions $X$ and $\Omega$.

\subsection{ Conditions for propagating Majorana eigenstates}
\label{subsection:propeig}

We have shown in subsection \ref{subsection:CPCTeig} that, as expected
since Majorana fermions are $C$ eigenstates,
a $C + PCT$ invariant propagator propagates Majorana fermions.

We now try to answer the reverse question {\em i.e.} which are the 
conditions on the propagator,
in particular concerning discrete symmetries, for it to propagate Majorana
fermions. This could look rather academic since we deal with one flavor and
that it is ``well known'' that, in particular, no $CP$ violating phase can
occur in this case. So, we ask the reader to consider this section as a
kind of  intellectual exercise. In addition to being a preparation to the
more complete study with several generations, it is also
motivated by the fact that, in the propagator formalism (which differs
from the one with  a classical Lagrangian endowed with a mass matrix),
even for one flavor,  a fermion and its antifermions get mixed as soon
as one allows all possible Lorentz invariant terms.
That this peculiarity can  {\em a priori} introduce
a mixing angle between a particle and its antiparticle (like for neutral
kaons) suggests that the
situation may not be so trivial as naively expected.
This section can also be considered as a test of the ``common sense''
statement  that, since Majorana fermions are defined as $C$ eigenstates,
 a propagator can only be expected to propagate Majorana fermions if it
satisfies the constraints cast by $C$ invariance.
We shall indeed reach a conclusion close to this one in the
following, with the only difference that $CP$ symmetry also enters the
game, for reasons that will be easy to understand
(the general demonstration for a number of flavors greater than
one, has been postponed to a further work).

\subsubsection{General conditions for diagonalizing a $\boldsymbol{PCT}$
invariant propagator}

We consider the most general $PCT$ invariant propagator (\ref{eq:propPCT}).

We are only concerned here with neutral fermions, for which diagonalizing
each $2 \times 2$ sub-matrix of the propagator is meaningful: for charged
fermions, this would mix in the same state fermions of different charges,
which is impossible as soon as we assume that electric charge is
conserved.

The two diagonal $2\times 2$ sub-blocks involve differential operators, with
one dotted an one undotted spinor index, factorized by simple functions of
space-time. We will suppose that, inside each of these sub-blocks, the four
differential operators are identical, such that their elements only differ
by the functions of space-time. When we speak about diagonalizing these
matrices, this concerns accordingly the space-time functions; then the
differential operators follow naturally.

The mass-like sub-blocks are diagonal in spinor indices and
involve only functions of space-time.

The propagator $\cal P$ writes
\begin{equation}
{\cal P}=\left(\begin{array}{cc}
|\ n_L> & |\ n_R>
\end{array}\right)
\left(\begin{array}{ccc}
K_1 & \vline & M_1 \cr
\hline
M_2 & \vline & K_2
\end{array}\right)
\left(\begin{array}{c}
<n_L\ | \cr <n_R\ |
\end{array}\right).
\end{equation}
$K_1$, $K_2$, $M_1$ and $M_2$ have {\em a priori}
 no special properties, are not hermitian nor symmetric.

There always exist $U_1$ and $U_2$, which have no reason to be unitary,
such that
\begin{equation}
U_1^{-1} K_1 U_1 = \Delta_1\ diagonal,\quad
U_2^{-1} K_2 U_2 = \Delta_2\ diagonal,
\label{eq:diag1}
\end{equation}
such that the propagator rewrites
\begin{eqnarray}
{\cal P}&=&\left(\begin{array}{cc}
|\ n_L> U_1 & |\ n_R> U_2
\end{array}\right)
\left(\begin{array}{ccc}
\Delta_1 & \vline & U_1^{-1}M_1U_2 \cr
\hline
U_2^{-1}M_2U_1 & \vline & \Delta_2
\end{array}\right)
\left(\begin{array}{c}
U_1^{-1} <n_L\ | \cr U_2^{-1} <n_R\ |
\end{array}\right)\cr
&=&\left(\begin{array}{cc}
|\ \mathfrak N_L>  & |\ \mathfrak N_R>
\end{array}\right)
\left(\begin{array}{ccc}
\Delta_1 & \vline & U_1^{-1}M_1U_2 \cr
\hline
U_2^{-1}M_2U_1 & \vline & \Delta_2
\end{array}\right)
\left(\begin{array}{c}
 <N_L\ | \cr  <N_R\ |
\end{array}\right),\cr
with &&<N_L\ |=U_1^{-1}<n_L\ |\;,\ <N_R\ |=U_2^{-1}<n_R\ |\;,
\ | \mathfrak N_L> = |\ n_L>U_1\;,\ |\ \mathfrak N_R> = |\ n_R>U_2\;.\cr
&&
\end{eqnarray}

The propagator can be diagonalized $\Leftrightarrow$
\begin{equation}
U_1^{-1}M_1U_2 = D_1\ diagonal,\quad U_2^{-1}M_2U_1 = D_2\ diagonal.
\label{eq:diag2}
\end{equation}
That $[D_1,D_2]=0$ entails in particular
\begin{equation}
U_1^{-1}M_1M_2 U_1 = D_1D_2\ diagonal = D_2D_1= U_2^{-1}M_2M_1 U_2,
\label{eq:diag3}
\end{equation}
which coincides with the commutation of $M_1$ and $M_2$ only when
$U_1=U_2$.

Since $[\Delta_1, D_1D_2]=0=[\Delta_2,D_1D_2]$, one also gets
$U_1^{-1}[K_1,M_1M_2]U_1 = 0 = U_2^{-1}[K_2,M_2M_1]U_2$, which entails
\begin{equation}
[K_1,M_1M_2] = 0 = [K_2,M_2M_1].
\label{eq:diag4}
\end{equation}
(\ref{eq:diag1}), (\ref{eq:diag2}), (\ref{eq:diag3}) and (\ref{eq:diag4})
are the conditions
that $K_1$, $K_2$, $M_1$ and $M_2$ must satisfy for the propagator to be
diagonalizable; they are must less stringent than the commutation of the
four of them.

{\bf In practice:}\quad
One supposes  that $M_1$ and $M_2$ fulfill condition (\ref{eq:diag4}).
To determine $U_1$ and $U_2$, one can  accordingly use indifferently
 (\ref{eq:diag1}) or (\ref{eq:diag3}):
$U_1$ diagonalizes $K_1$ or $M_1M_2$, $U_2$ diagonalizes
$K_2$ or $M_2M_1$.  Supposing that  (\ref{eq:diag3}) is satisfied,
$M_1 M_2$ and of $M_2 M_1$ are constrained to have the same eigenvalues,
which may give additional restrictions on $M_1$ and $M_2$.

Once $U_1$ and $U_2$ are determined,  call
\begin{equation}
{\cal M}_1 = U_1^{-1}M_1 U_2, \quad{\cal M}_2 = U_2^{-1}M_2 U_1.
\end{equation}
(\ref{eq:diag3}) entails that, in particular, ${\cal M}_1$ and ${\cal M}_2$
must commute.
Since $U_1$ diagonalizes
$M_1M_2$ and $U_2$ diagonalizes $M_2M_1$, 
 ${\cal M}_1{\cal M}_2$ and ${\cal M}_2{\cal M}_1$ are diagonal.

Write ${\cal M}_1= \left(\begin{array}{cc} \mathfrak{a} & \mathfrak{b} \cr
\mathfrak{c} & \mathfrak{d}
\end{array}\right)$ and
${\cal M}_2= \left(\begin{array}{cc} \mathfrak{p} & \mathfrak{q} \cr
\mathfrak{r} & \mathfrak{s}
\end{array}\right)$; by direct inspection, one finds that the two products
${\cal M}_1 {\cal M}_2$ and ${\cal M}_2 {\cal M}_1$
are diagonal either if ${\cal M}_1$ and ${\cal M}_2$ are diagonal, or if
${\cal M}_2= t\left(\begin{array}{rr} \mathfrak{d} & -\mathfrak{b} \cr
-\mathfrak{c} & \mathfrak{a}
\end{array}\right)$, that is, is proportional to ${\cal M}_1^{-1}$;
in this last case,
${\cal M}_1{\cal M}_2 = {\cal M}_2{\cal M}_1$ is proportional to the unit
matrix, which means that the eigenvalues of $M_1M_2$ are all identical
 (and so are the eigenvalues of $M_2M_1$).

We are looking for more: the conditions that must satisfy $M_1$ and $M_2$ 
for ${\cal M}_1$ and ${\cal M}_2$ to be separately diagonal. We attempt to
find them by putting the additional restriction that the eigenstates are
Majorana fermions.

\subsubsection{Condition for propagating Majorana fermions}

A necessary (but not sufficient) condition for the propagating states to
be Majorana is that, by some change of basis, the
propagator can be cast in the form
\begin{equation}
\Delta_{Maj}(p^2) =
\left(\begin{array}{ccccc}
\left(\begin{array}{cc}a_1(p^2) & \cr
  & b_1(p^2)\end{array}\right)p_\mu\sigma^\mu_{\alpha\beta}
 & \vline &
\left(\begin{array}{cc}m_1(p^2) &  \cr
 & \mu_1(p^2)\end{array}\right)\delta_{\alpha\beta} \cr
\hline
\left(\begin{array}{cc}m_2(p^2) & \cr
 & \mu_2(p^2)\end{array}\right)\delta_{\alpha\beta}
& \vline &
\left(\begin{array}{cc}a_2(p^2) &  \cr
  & b_2(p^2)\end{array}\right)p_\mu\overline{\sigma^\mu}_{\alpha\beta}
\end{array}\right),
\label{eq:majprop}
\end{equation}
with four diagonal $2\times 2$ sub-blocks. Indeed, on can then decompose
the propagator into two $4\times 4$ propagators (in a shortened
notation)
$\left(\begin{array}{cc} a_1 & m_1 \cr m_2 & a_2\end{array}\right)$ and
$\left(\begin{array}{cc} b_1 & \mu_1 \cr \mu_2 & b_2\end{array}\right)$,
and the Majorana fermions (see subsection \ref{subsection:majorana})
 are  eventually, respectively, composed with the
first components of $n_L$ and $n_R$, and with the second components of the
same set.
So, in particular, both  kinetic-like  and mass-like terms, should be
diagonalizable simultaneously
\footnote{Imposing commutation relations between all $2\times 2$ sub-blocks
of the propagator is excessive.}
.
We note
\begin{eqnarray}
U_1^{-1} = \left(\begin{array}{cc} a & b \cr c & d \end{array}\right),
\quad
U_2^{-1} = \left(\begin{array}{cc} p & q \cr r & s \end{array}\right),
\quad
D_1=\left(\begin{array}{cc} d_1 & 0 \cr 0 & \delta_1 \end{array}\right),
\quad
D_2=\left(\begin{array}{cc} d_2 & 0 \cr 0 & \delta_2 \end{array}\right).
\label{eq:defD}
\end{eqnarray}
One has
\begin{eqnarray}
&&<N_L\ | = \left(\begin{array}{c}a<\xi^\alpha\ | +
b<(-i)(\eta^{\dot\alpha})^\ast\ | \cr
      c<\xi^\alpha\ | + d<(-i)(\eta^{\dot\alpha})^\ast\
|\end{array}\right),
\cr
&&<N_R\ | = \left(\begin{array}{c}p<(-i)\xi_\alpha^\ast\ | + q
<\eta_{\dot\alpha}\ | \cr
      r<(-i)\xi_\alpha^\ast\ | + s <\eta_{\dot\alpha}\
|\end{array}\right),
\cr
&&|\ \mathfrak{N}_L> = \frac{1}{ad-bc}
\left(\begin{array}{cc} d |\ \xi^\alpha >-c |\ (-i)(\eta^{\dot\alpha})^\ast> &
   -b |\ \xi^\alpha>  +a |\ (-i)(\eta^{\dot\alpha})^\ast>\end{array}\right),
\cr
&&|\ \mathfrak{N}_R> = \frac{1}{ps-qr}
\left(\begin{array}{cc} s |\ (-i)\xi_\alpha^\ast> -r |\ \eta_{\dot\alpha}> &
   -q |\ (-i)\xi_\alpha^\ast>  + p |\
\eta_{\dot\alpha}>\end{array}\right),\cr
&&
\label{eq:eigmaj}
\end{eqnarray}
and the question is whether the propagator $<0\ |\ {\cal T}\left(\begin{array}{c}
N_L (x)\cr N_R (x)
\end{array}\right) \left(\begin{array}{cc} {\mathfrak N_L}(-x) & {\mathfrak
N_R}(-x) \end{array}\right)^\dagger\ |\ 0>$ can be identified with that
of  a Majorana fermion and its antifermion (that is, itself) .
Eq.~(\ref{eq:eigmaj}) yields in particular the four mass-like propagators
\begin{eqnarray}
&&<0\ |{\cal T} \left(d \xi^\alpha +ic (\eta^{\dot\alpha})^\dagger\right) (x)
 \left(ip^\ast \xi_\beta +q^\ast  (\eta_{\dot\beta})^\dagger\right) (-x)|\ 0> =
(ad-bc) d_1(x)\delta_{\alpha\beta},\quad (a)\cr
&&<0\ |{\cal T} \left(-b \xi^\alpha -ia (\eta^{\dot\alpha})^\dagger\right) (x)
 \left(ir^\ast \xi_\beta +s^\ast  (\eta_{\dot\beta})^\dagger\right) (-x)|\ 0> =
(ad-bc) \delta_1(x)\delta_{\alpha\beta},\quad (b)\cr
&&<0\ |{\cal T} \left(-is (\xi_\alpha)^\dagger -r \eta_{\dot\alpha}\right) (x)
 \left(a^\ast (\xi^\beta)^\dagger +ib^\ast  \eta^{\dot\beta}\right) (-x)|\ 0> =
(ps-qr) d_2(x)\delta_{\alpha\beta},\quad (c)\cr
&&<0\ |{\cal T} \left(iq (\xi_\alpha)^\dagger +p \eta_{\dot\alpha}\right) (x)
 \left(c^\ast (\xi^\beta)^\dagger +id^\ast  \eta^{\dot\beta}\right) (-x)|\ 0> =
(ps-qr) \delta_2(x)\delta_{\alpha\beta},\quad (d)\cr
&&
\label{eq:mprop1}
\end{eqnarray}
which must be the only four non-vanishing such propagators since $U_1^{-1}M_1
U_2$ and $U_2^{-1} M_2 U_1$ must be diagonal.
We have to identify them with typical mass-like Majorana propagators.
For that purpose, 
we have a priori to introduce two Majorana fermions; $X_M^\pm
=\left(\begin{array}{c} \zeta^\alpha \cr \pm (-i)
(\zeta_\alpha)^\ast\end{array}\right)$,
associated, together with its antifermion, to $(N_L, N_R)$,
and $Y_M^\pm =\left(\begin{array}{c} \chi^\beta \cr
\pm (-i) (\chi_\beta)^\ast\end{array}\right)$, associated,
together with its antifermion,
to $({\mathfrak N}_L, {\mathfrak N}_R)$.  An $X-Y$ propagator
\footnote{We allow here  $X \not = Y$, but will then become  more restrictive
by requesting $X = Y$, which better corresponds  to the intuitive
picture of propagating a definite Majorana fermion.}
reads (we go to the $\overline{( )}$ fields, which introduces
an extra $\gamma^0$;
this has in particular for consequence that ``mass-like''
propagators now appear on the diagonal)
\begin{equation}
\hskip -.5cm<0\ |{\cal T} X_M(x)  \overline{Y_M}(-x)|\ 0>
=
\left(\begin{array}{cc}
<0\ |{\cal T} \zeta^\alpha(x)  (\pm i)\chi_\beta(-x)|\ 0> &
<0\ |{\cal T} \zeta^\alpha(x)  (\chi^\beta)^\dagger(-x)|\ 0> \cr
<0\ | {\cal T}  (\zeta_\alpha)^\dagger (x) \chi_\beta(-x)|\ 0> &
<0\ | {\cal T}  (\mp i)(\zeta_\alpha)^\dagger (x) (\chi^\beta)^\dagger(-x)|\ 0>
\end{array}\right).
\label{eq:mprop2}
\end{equation}
The four lines of (\ref{eq:mprop1}) correspond to two mass-like
$X-Y$ propagators  only if one can associate them into two pairs, such
that each pair has the same structure as the  diagonal terms
of (\ref{eq:mprop2}).
There are accordingly two possibilities:  pairing  (a) with (c)
and (b) with (d), or (a) with (d) and (b) with (c).

\smallskip

$\ast$\ \underline{The first possibility} requires
($\kappa$ and $\lambda$ are proportionality constants)
$p = i\lambda a^\ast, q = i\lambda
b^\ast, r=-i \kappa c^\ast, s = -i\kappa d^\ast$, such that
\begin{equation}
U_2^{-1}= i\left(\begin{array}{rr} \lambda a^\ast & \lambda b^\ast \cr
-\kappa c^\ast &  -\kappa d^\ast \end{array}\right). 
\label{eq:mpcond1}
\end{equation}

$\ast$\  \underline{The second possibility} requires
$p=i\rho c^\ast, q=i\rho d^\ast,r=i\theta a^\ast,  s=i\theta b^\ast$
such that
\begin{equation}
U_2^{-1}= i\left(\begin{array}{rr} \rho c^\ast & \rho d^\ast \cr
\theta a^\ast &  \theta b^\ast \end{array}\right). 
\label{eq:mpcond2}
\end{equation}

\medskip

{\it From now onwards, we furthermore request that a single Majorana fermion
propagates} in the sense that only ${\cal T}$-products of the type
$<0\ |\ {\cal T} X^\alpha (x) X_\alpha(-x)\ |\ 0>$ occur, which associates
$|\ {\cal N}_L> = |\ X^\alpha>$ and $< N_R\ | = <X_\alpha^\ast\ |$.
The only possibility is that the coefficients of $|\ {\cal N}_L>$ and
$|\ N_R>$ in (\ref{eq:eigmaj}) be proportional, and so be the ones of
$|\ {\cal N}_R>$ and $|\ N_L>$ (the two sets of conditions are the same);
this gives the supplementary conditions ($\sigma$ and $\beta$ are two
other proportionality constants) 
$ p = i\sigma d^\ast, q=-i\sigma c^\ast, r= -i\gamma b^\ast,
s= i\gamma a^\ast
$, such that
\begin{equation}
U_2^{-1}= i\left(\begin{array}{rr} \sigma d^\ast & -\sigma c^\ast \cr
-\gamma b^\ast &  \gamma a^\ast \end{array}\right). 
\label{eq:ideigen}
\end{equation}
* \underline{First possibility} ($U_2^{-1}$ is given by (\ref{eq:mpcond1}) above).

Compatibility between (\ref{eq:mpcond1}) and (\ref{eq:ideigen}) requires
$\frac{q}{p} = \frac{b^\ast}{a^\ast}=-\frac{c^\ast}{d^\ast} =
-\frac{r}{s} = \omega^\ast$ such that we end up with
\begin{equation}
U_1^{-1}= \left(\begin{array}{cc} a & \omega a \cr -\omega d &
d\end{array}\right),\quad
U_2^{-1}= \left(\begin{array}{cc} p & \omega^\ast p \cr -\omega^\ast s &
s\end{array}\right)
= \left(\begin{array}{cc}
\lambda a^\ast & \lambda \omega^\ast a^\ast \cr
\kappa \omega^\ast d^\ast & -\kappa d^\ast \end{array}\right).
\label{eq:mpcond3}
\end{equation}

We look for $PCT$ invariant 
$M_1= \left(\begin{array}{cc} m_{L1}(x) & \mu_1(x) \cr
                     \mu_1(x) & m_{R1}(x) \end{array}\right)$ and
$M_2= \left(\begin{array}{cc} m_{L2}(x) & \mu_2(x) \cr
                     \mu_2(x) & m_{R2}(x) \end{array}\right)$
(see (\ref{eq:propPCT}))
and their diagonalization according to (\ref{eq:diag2}) and (\ref{eq:defD})
by $U_1$ and $U_2$ given by (\ref{eq:mpcond3}) and satisfying
(\ref{eq:diag4}).

The equations (\ref{eq:diag1}) of diagonalization
for the kinetic-like terms $K_1 = \left(\begin{array}{cc} \alpha & u
\cr v & \beta \end{array}\right)$ and
$K_2= \left(\begin{array}{cc} \alpha & v
\cr u & \beta \end{array}\right)$ (see (\ref{eq:propPCT}))
yield,  for the vanishing of the non-diagonal terms, the conditions
\begin{eqnarray}
u - \omega^2 v &=& \omega(\alpha-\beta),\cr
v - \omega^2 u &=& \omega(\alpha -\beta),\cr
v -\omega^{\ast 2} u &=& \omega^\ast(\alpha-\beta),\cr
u - \omega^{\ast 2} v &=& \omega^\ast(\alpha-\beta).
\label{eq:kincon}
\end{eqnarray}
Likewise, the diagonalization equations (\ref{eq:diag2})  for the mass-like terms
yield
\begin{eqnarray}
\omega^\ast m_{L1} -\omega m_{R1} &=& \mu_1(1-|\omega|^2),\cr
\omega m_{L1} -\omega^\ast m_{R1} &=& \mu_1(1-|\omega|^2),\cr
\omega^\ast m_{L2} -\omega m_{R2} &=& \mu_2(1-|\omega|^2),\cr
\omega m_{L2} -\omega^\ast m_{R2} &=& \mu_2(1-|\omega|^2).
\label{eq:mcon}
\end{eqnarray}
First, we eliminate the trivial case $\omega=1$ which brings back to a $C$
invariant propagator.

Subtracting the first or the last two equations of (\ref{eq:kincon})
yields $u=v$.
One then gets $\alpha-\beta = u\frac{1-\omega^2}{\omega} =
u\frac{1-\omega^{\ast 2}}{\omega^\ast}$, such that $\omega$ must be real.

Subtracting the first two equations of (\ref{eq:mcon}) also shows that
$\omega$ must be real as soon as one supposes $m_{L1}+m_{R1} \not=0$, which
we do. Then, one gets 
$\frac{\mu_1}{m_{L1}-m_{R1}} = \frac{\omega}{1-\omega^2}
= \frac{\mu_2}{m_{L2}-m_{R2}}$. Gathering the results from
(\ref{eq:kincon}) and (\ref{eq:mcon}) leads accordingly to
\begin{eqnarray}
 K_1 &=& u\left(\begin{array}{cc} \alpha &
(\alpha-\beta)\displaystyle\frac{\omega}{1-\omega^2} \cr
(\alpha-\beta)\displaystyle\frac{\omega}{1-\omega^2} &
\beta\end{array}\right)=K_2,\cr
M_1 &=& \left(\begin{array}{cc} m_{L1} &
(m_{L1}-m_{R1})\displaystyle{\frac{\omega}{1-\omega^2}} \cr
(m_{L1}-m_{R1})\displaystyle{\frac{\omega}{1-\omega^2}} & m_{R1}\end{array}\right),\cr
M_2 &=& \left(\begin{array}{cc} m_{L2} &
(m_{L2}-m_{R2})\displaystyle{\frac{\omega}{1-\omega^2}} \cr
(m_{L2}-m_{R2})\displaystyle{\frac{\omega}{1-\omega^2}} & m_{R1}\end{array}\right),
\end{eqnarray}
and we shall hereafter write $\omega = \tan\vartheta$.
 The four real symmetric matrices $K_1 = K_2, M_1, M_2$ can be
simultaneously diagonalized by the same rotation matrix $U(\vartheta)$
 of angle $\vartheta$.
After diagonalization, the propagator writes
\begin{eqnarray}
\Delta &=&
\left( \begin{array}{cc} |\ n_L> U & |\ n_R> U \end{array}\right)
\left(\begin{array}{ccccc}
\delta_+ & & \vline & \mu_{1+} & \cr
& \delta_- & \vline & & \mu_{1-} \cr
\hline
\mu_{2+} & & \vline & \delta_+ & \cr
& \mu_{2-} & \vline & & \delta_- \end{array}\right)
\left(\begin{array}{c} U^T <n_L\ | \cr U^T <n_R\ |
\end{array}\right),\cr
\text{with}&&\ \delta_\pm =\frac12\left(\alpha+\beta
\pm\frac{\alpha-\beta}{\cos 2\vartheta}\right),\quad
\mu_{1,2,\pm} =\frac12\left(m_{L1,2} + m_{R1,2}
\pm\frac{m_{L1,2} -m_{R1,2}}{\cos 2\vartheta}\right).
\end{eqnarray}
To propagate a Majorana fermion, the condition $\mu_{1+} = \mu_{2+}$
should furthermore be fulfilled. This requires, for arbitrary $\vartheta$,
 $m_{R1} = m_{R2}, m_{L1}=m_{L2}$ (and thus $\mu_1 = \mu_2$).
This corresponds to a propagator (before diagonalization)
\begin{equation}
\Delta =
\left( \begin{array}{cc} |\ n_L>  & |\ n_R>  \end{array}\right)
\left(\begin{array}{ccccc}
\alpha &u & \vline & m_L & \mu \cr
u & \beta & \vline & \mu & m_R \cr
\hline
m_L & \mu & \vline & \alpha & u \cr
\mu & m_R & \vline &u &\beta  \end{array}\right)
\left(\begin{array}{c}  <n_L\ | \cr  <n_R\ |
\end{array}\right),\ \frac{u}{\alpha-\beta}=\frac{\mu}{m_L-m_R},
\end{equation}
that is, a  $CP$ invariant propagator (see (\ref{eq:CPPCTprop2}))
(the $C$ invariant case corresponds to
$\omega =1$ (see (\ref{eq:CPCTprop1})), which has been treated previously).
The propagating  Majorana fermion are \newline
$\psi_M=\left(\begin{array}{c}
\cos\vartheta\xi^\alpha -\sin\vartheta (-i (\eta^{\dot\alpha})^\ast) \cr
\cos\vartheta (-i (\xi_\gamma)^\ast) -\sin\vartheta \eta_{\dot\gamma}
\end{array}\right)$ and
$\chi_M = \left(\begin{array}{c}
\sin\vartheta \xi^\alpha + \cos\vartheta(-i (\eta^{\dot\beta})^\ast \cr
\sin\vartheta   (-i (\xi_\gamma)^\ast) + \cos\vartheta\eta_{\dot\gamma} \cr
\end{array}\right)$. 
\medskip

*\  \underline{Second possibility} ($U_2^{-1}$ is given by (\ref{eq:mpcond2}) above).
Equating (\ref{eq:ideigen}), (\ref{eq:mpcond2}) and  the expression for
$U_2^{-1}$ in (\ref{eq:defD}), one gets $q/p = d^\ast/c^\ast =
-c^\ast/d^\ast$, $s/r = b^\ast/a^\ast = -a^\ast/b^\ast$, which gives
$d=\pm ic, b=\pm ia$ and thus
\begin{equation}
U_1^{-1} = \left(\begin{array}{cc} a & \pm ia \cr
                                   c & \pm ic \end{array}\right),\quad
U_2^{-1}= i\left(\begin{array}{cc} \rho c^\ast & \mp i \rho c^\ast \cr
                        \pm i \gamma a^\ast & \gamma a^\ast
\end{array}\right).
\end{equation}
The diagonalization equations (\ref{eq:diag2})  for the mass-like terms
yield, for the vanishing of the non-diagonal terms, the conditions
\begin{eqnarray}
m_{L1} &=& -m_{R1},\cr
m_{L2} &=& -m_{R2}.
\end{eqnarray}
The equations (\ref{eq:diag1}) of diagonalization
for the kinetic-like terms
yield the conditions
\begin{eqnarray}
u + v &=& \pm i(\alpha-\beta),\cr
u + v &=& \pm i(\beta-\alpha),
\label{eq:kincon2}
\end{eqnarray}
which require $v=-u, \beta =\alpha$.

So, the kinetic and mass-like propagators write
\begin{eqnarray}
K_1 = \left(\begin{array}{rr} \alpha & u \cr -u & \alpha\end{array}\right),
&\quad&
K_2 = \left(\begin{array}{rr} \alpha & -u \cr u & \alpha\end{array}\right),
\cr
M_1 = \left(\begin{array}{rr} m_1 & \mu_1 \cr \mu_1 &
-m_1\end{array}\right),
&\quad&
M_2 = \left(\begin{array}{rr} m_2 & \mu_2 \cr \mu_2 &
-m_2\end{array}\right).
\end{eqnarray}
$K_1$ and $K_2$, which commute, can be diagonalized simultaneously by a
single matrix $U$.
The conditions (\ref{eq:diag4}) $[K_1, M_1M_2]=0= [K_2,M_2M_1]$ require
$m_1/m_2 = \mu_1/\mu_2$, such that  $M_2 = \chi M_1$. Since $U_1=U=U_2$,
the diagonalization equations (\ref{eq:diag2}) for the mass-like propagators
rewrite $U^{-1} M_1 U = D_1, U^{-1} M_2 U = \chi D_1$, such that the set of
four matrices $K_1, K_2, M_1, M_2$ must commute, which requires $u=0$. The
kinetic-like propagators are thus ``standard'', {\em i.e.} proportional to
the unit matrix.
Before diagonalization, the propagator writes
\begin{eqnarray}
\Delta &=&
\left( \begin{array}{rr} |\ n_L>  & |\ n_R>  \end{array}\right)
\left(\begin{array}{ccccc}
\alpha & & \vline & m_1 & \mu_1 \cr
& \alpha & \vline & \mu_1 & -m_1 \cr
\hline
\chi m_1 & \chi \mu_1 & \vline & \alpha & \cr
\chi \mu_1 & -\chi m_1 & \vline & & \alpha \end{array}\right)
\left(\begin{array}{c}  <n_L\ | \cr  <n_R\ |
\end{array}\right),
\label{eq:pr2}
\end{eqnarray}
and, after diagonalization, 
\begin{eqnarray}
\Delta &=&
\left( \begin{array}{rr} |\ n_L> U & |\ n_R> U \end{array}\right)
\left(\begin{array}{ccccc}
\alpha & & \vline & \mu & \cr
& \alpha & \vline & & -\mu \cr
\hline
\chi\mu & & \vline & \alpha & \cr
& -\chi\mu & \vline & & \alpha \end{array}\right)
\left(\begin{array}{c} U^T <n_L\ | \cr U^T <n_R\ |
\end{array}\right),\cr
&&\text{with}\quad \mu = \sqrt{m_1^2 + \mu_1^2}.
\label{eq:pr3}
\end{eqnarray}
It can propagate Majorana fermions only if $\chi=1$, such that $M_1 = M_2$.
Then, (\ref{eq:pr2}) is a special kind of $PC$ invariant propagator (see
(\ref{eq:CPPCTprop2})), which becomes $C$ invariant only when $m_1=0$.
The two Majorana fermions have masses  $\pm\mu/\alpha$. They are 
$\psi_M=\left(\begin{array}{c}
\cos\vartheta\xi^\alpha -\sin\vartheta (-i (\eta^{\dot\alpha})^\ast) \cr
\cos\vartheta (-i (\xi_\gamma)^\ast) -\sin\vartheta \eta_{\dot\gamma}
\end{array}\right)$ and
$\chi_M = \left(\begin{array}{c}
\sin\vartheta \xi^\alpha + \cos\vartheta(-i (\eta^{\dot\beta})^\ast \cr
\sin\vartheta   (-i (\xi_\gamma)^\ast) + \cos\vartheta\eta_{\dot\gamma} \cr
\end{array}\right)$, 
with $\tan 2\vartheta = \mu_1/m_1$.

\subsubsection{Conclusion}

For one flavor (particle + antiparticle),
a necessary condition for the eigenstates of the propagator
to be Majorana is either that this propagator (supposed
 to satisfy the
constraints cast by $PCT$ invariance) satisfies the constraints cast by
 $C$ invariance (which corresponds to $\omega=1$) or by
$CP$ invariance
\footnote{This is linked to the property of Majorana fermions to
 have $CP$ {\em parity} $=\pm i$ (see subsections \ref{subsection:CPtrans} and
\ref{subsection:majorana}). The two corresponding $(\pm i \gamma^0)$ factors
cancel in the $\cal T$-product of their propagator, which makes it
$CP$ invariant. This explains why not only $C$ invariant, but also
$CP$ invariant propagators can propagate
Majorana fermions}. So,  reciprocally, 
if the most general $PCT$ invariant propagator for one flavor
does not satisfy the constraints cast by $C$ nor the ones cast by $CP$,
its eigenstates cannot be Majorana.

\section{General conclusion}
\label{section:conclusion}

In this work, we have extended the propagator approach 
\cite{JacobSachs} \cite{Sachs} \cite{MaNoVy} to coupled fermionic
systems.  It is motivated, in particular, by the ambiguities  that
unavoidably occur when dealing with a classical fermionic Lagrangian
endowed with a mass matrix.
The goal of this formalism is, in particular, to determine at
which condition the propagating neutral fermions, defined as the
eigenstates, at the poles, of their full propagator, are Majorana.
Due to the intricacies of this approach, we presently
limited ourselves to the simplest case of a single fermion
and its antifermion. Since Lorentz invariance allows that they
get coupled (as long as it is not forbidden by electric charge
conservation), one can expect properties similar to the ones of
the neutral kaons system.  In this simple
case, we have proved what is suggested by common sense, {\em i.e.}
that the propagating fermions can only be Majorana if their propagator
satisfies the constraints cast by $C$ (or $CP$) invariance.

The generalization to several flavors  will be the object of a
subsequent work, with, in particular, the persistent goal of unraveling
the  nature of neutrinos.

\vskip .5cm
\begin{em}
\underline {Acknowledgments}: conversations, comments and critics with /
from V.A.~Novikov,  M.I.~Vysotsky and J.B.~Zuber are gratefully acknowledged.
\end{em}
%

\newpage\null

\appendix

\section{Notations: spinors}
\label{section:nota}

\subsection{Weyl spinors}
\label{subsection:spinors}

We adopt the notations of \cite{Landau}, with undotted and dotted indices.

Undotted spinors, contravariant $\xi^\alpha$ or covariant
$\xi_\alpha$ can be also called left spinors.
Dotted spinors, covariant $\eta_{\dot\alpha}$ or contravariant
$\eta^{\dot\alpha}$ can then be identified as right spinors.
They are  2-components complex spinors. The 2-valued spinor indices are not
explicitly written.

By an arbitrary transformation of the proper Lorentz group
\begin{equation}
\alpha\delta - \beta\gamma = 1,
\end{equation}
they transform by
\begin{eqnarray}
\xi^{1'} &=& \alpha \xi^1 + \beta \xi^2,\cr
\xi^{2'} &=& \gamma \xi^1 + \delta \xi^2,\cr
&&\cr
\eta^{\dot 1'} &=& \alpha^\ast \eta^{\dot 1}+ \beta^\ast \eta^{\dot 2},\cr
\eta^{\dot 2'} &=& \gamma^\ast \eta^{\dot 1}+ \delta^\ast \eta^{\dot 2}.
\end{eqnarray}
To raise or lower spinor indices, one has to use the metric of $SL(2,C)$
\begin{equation}
g_{\alpha\beta} = \left(\begin{array}{rr} 0 & 1 \cr
                                          -1 & 0 \end{array}\right) =
i\sigma^2_{\alpha\beta};\quad
g^{\alpha\beta}= \left(\begin{array}{rr} 0 & -1 \cr
                                           1 & 0 \end{array}\right) =
-i(\sigma^2)_{\alpha\beta},
\end{equation}
and the same for dotted indices. The $\sigma^2$ matrix will always be
represented with indices down.
\begin{equation}
\xi_\alpha = g_{\alpha\beta} \xi^\beta =i\sigma^2_{\alpha\beta}\xi^\beta,
\eta^{\dot\alpha}= g^{\dot\alpha\dot\beta}\eta_{\dot\beta}
=-i\sigma^2_{\dot\alpha\dot\beta}\eta_{\dot\beta}.
\end{equation}
One has
\begin{equation}
\xi.\zeta =\xi^\alpha \zeta_\alpha = \xi^1\zeta^2-\xi^2\zeta^1 =
-\xi_\alpha\zeta^\alpha\ invariant.
\end{equation}
By definition, $\eta_{\dot\alpha} \sim {\xi_\alpha}^\ast$ (transforms
as);
\begin{equation}
\eta_{\dot\alpha} \sim (g_{\alpha\beta} \xi^\beta)^\ast
= g_{\alpha\beta} (\xi^\beta)^\ast =i\sigma^2_{\alpha\beta} \xi^{\beta\ast}:
\end{equation}
a right-handed Weyl spinor and
the complex conjugate
of a left-handed Weyl spinor transform alike by Lorentz;
likewise, a left-handed spinor transforms like
the complex conjugate of a right-handed spinor.

A Dirac (bi-)spinor is
\begin{equation}
\xi_D = \left(\begin{array}{c} \xi^\alpha \cr
\eta_{\dot\alpha}\end{array}\right).
\label{eq:defDirac}
\end{equation}

\subsection{Pauli and Dirac matrices}
\label{subsection:PD}

Since we work with Weyl fermions, we naturally choose the Weyl representation.

Pauli matrices:
\begin{equation}
\sigma^0 = \left( \begin{array}{rr}  1 & 0 \cr
                                     0 & 1 \end{array}\right),
\sigma^1 = \left( \begin{array}{rr}  0 & 1 \cr
                                     1 & 0 \end{array}\right),
\sigma^2 = \left( \begin{array}{rr}  0 & -i \cr
                                     i & 0 \end{array}\right),
\sigma^3 = \left( \begin{array}{rr}  1 & 0 \cr
                                     0 & -1 \end{array}\right);
\label{eq:pauli1}
\end{equation}
$\gamma$ matrices
\begin{equation}
\gamma^0 = \left( \begin{array}{rrcrr}  0 & 0 & \vline & 1 & 0 \cr
                                       0 & 0 & \vline & 0 & 1 \cr
                                              \hline
                                       1 & 0 & \vline & 0 & 0 \cr
                                       0 & 1 & \vline & 0 & 0 \end{array}\right),
\gamma^i =  \left( \begin{array}{rr}  0 & -\sigma^i \cr
                                     \sigma^i & 0 \end{array}\right),
\gamma_5 = i\gamma^0\gamma^1\gamma^2\gamma^3 =
\left( \begin{array}{rrcrr}   1 & 0 & \vline & 0 & 0 \cr
                             0 & 1 & \vline & 0 & 0 \cr
                                        \hline
                             0 & 0 & \vline & -1 & 0 \cr
                             0 & 0 & \vline & 0 & -1 \end{array}\right),
\label{eq:gamma1}
\end{equation}
and one notes
\begin{equation}
\gamma^\mu = (\gamma^0, \vec\gamma) = \gamma^0
\left( \begin{array}{rr} {\sigma^\mu} & 0 \cr
                             0      & \overline{\sigma^\mu}\end{array}\right),
\label{eq:gamma2}
\end{equation}
with
\begin{equation}
\sigma^\mu = (\sigma^0, \vec\sigma),\quad \overline{\sigma^\mu} = (\sigma^0,
-\vec\sigma),\quad \vec\sigma=(\sigma^1,\sigma^2,\sigma^3).
\label{eq:pauli3}
\end{equation}
\begin{eqnarray}
&& (\gamma^0)^\dagger = \gamma^0, (\gamma^5)^\dagger = \gamma^5,
(\gamma^{1,2,3})^\dagger = -\gamma^{1,2,3},\cr
&& (\gamma^0)^\ast = \gamma^0, (\gamma^5)^\ast = \gamma^5,
(\gamma^{1,3})^\ast = \gamma^{1,3}, (\gamma^2)^\ast = -\gamma^2,\cr
&& (\gamma^0)^2=1,(\gamma^5)^2=1,(\gamma^{1,2,3})^2=-1,\cr
&& \gamma^0(\gamma^0)^\dagger =1,\gamma^5(\gamma^5)^\dagger =1,
\gamma^{1,2,3}(\gamma^{1,2,3})^\dagger =1.
\end{eqnarray}
One has
\begin{equation}
(\sigma^0)^2 = 1 = (\sigma^i)^2, \{\sigma^i,\sigma^j\} = 2\delta^{ij}.
\label{eq:pauli2}
\end{equation}
One has the relation
\begin{equation}
\sigma^2_{\beta\delta}\sigma^2_{\alpha\gamma}
= \delta_{\beta\gamma}\delta_{\alpha\delta} -
\delta_{\alpha\beta}\delta_{\delta\gamma},
\label{eq:sigmaprod}
\end{equation}
and the following one is very useful
\begin{equation}
\sigma^2 \sigma^i \sigma^2 = -(\sigma^i)^\ast,\quad \sigma^2 \sigma^0
\sigma^2 = \sigma^0 \Rightarrow
\sigma^2 \sigma^\mu\sigma^2 = (\sigma^0, -\vec \sigma^\ast) =
\overline{\sigma^\mu}^\ast.
\label{eq:sigmarel}
\end{equation}
As far as kinetic terms are concerned,
\begin{equation}
\gamma^0 \gamma^\mu p_\mu = (\gamma^0)^2 p_\mu
\left( \begin{array}{rr} {\sigma^\mu} & 0 \cr
                             0      & \overline{\sigma^\mu}\end{array}\right)
=\left(\begin{array}{cc} p^0 - \vec p. \vec\sigma & 0 \cr
                             0      & p^0 + \vec p.\vec\sigma\end{array}\right)
.
\label{eq:kin}
\end{equation}
%

\section{The adjoint of an antilinear operator}
\label{section:adjointanti}

Following Weinberg  \cite{Weinberg},
 let us show that the adjoint of an antilinear operator (see
(\ref{eq:antilin}) for the definition)
$\cal A$ cannot be defined by $<{\cal A}\psi\ |\ \chi> = <\psi\ |\
{\cal A}^\dagger\ |\ \chi>$
\footnote{This changes nothing to our demonstrations.}
.
Indeed, suppose that we can take the usual definition above, and
let $c$ be a c-number; using the antilinearity of $\cal A$ one gets
$<{\cal A}(c\psi) \ |\ \chi> = <c^\ast ({\cal A}\psi)\ |\ \chi>
= c < ({\cal A}\psi)\ |\ \chi>=
 c<\psi\ |\ {\cal A}^\dagger\ |\ \chi>$ is linear in $\psi$.

But one has also $<{\cal A}(c\psi) \ |\ \chi> =<(c\psi)\ |\ {\cal
A}^\dagger\  |\ \chi> = <\psi\ |\ c^\ast{\cal A}^\dagger\ |\
\chi> = c^\ast <\psi\ |\ {\cal A}^\dagger\ |\ \chi>$ is antilinear in $\psi$,
which is incompatible with the result above. So, the two expressions cannot
be identical and $<{\cal A}\psi\ |\ \chi> \not = <\psi\ |\
{\cal A}^\dagger\ |\ \chi>$.

Weinberg (\cite{Weinberg} p.51) defines the adjoint by
\footnote{So defined, taking $\psi=\chi$, the adjoint satisfies
$<\psi\ |\ {\cal A}\ |\ \psi> = <\psi\ |\ {\cal A}^\dagger\ |\ \psi>$.
This entails in particular that, for a antiunitary operator 
\begin{equation}
<\psi\ |\ {\cal A}^\dagger\ |\ \psi>^\ast
\not= <\psi\ |\ {\cal A}\ |\ \psi>,
\label{attention}
\end{equation}
unless what happens for antiunitary operators (otherwise the matrix element
$<\psi \ |\ {\cal A}\ |\ \psi>$ of any antiunitary operator
could only be real, which is nonsense).
}
\begin{equation}
<\psi \ |\ {\cal A}^\dagger\ |\ \chi>
\equiv <\psi \ |\ {\cal A}^\dagger\ \chi> 
=< {\cal A} \psi\ |\ \chi>^\ast
= <\chi\ |\ {\cal A}\ \psi>
\equiv <\chi\ |\ {\cal A}\ |\ \psi>
\label{eq:adjoint}
\end{equation}
Then, even for an antilinear and antiunitary operator one has
\footnote{This is in contradiction with \cite{BrancoLavouraSilva}.}
\begin{equation}
{\cal A}^\dagger{\cal A}= 1.
\end{equation}
Indeed, $<\psi\ |\ {\cal A}^\dagger{\cal A}\ |\ \chi> =
<\psi\ |\ {\cal A}^\dagger\ |\ {\cal A} \chi> \stackrel{(\ref{eq:adjoint})}{=}
<{\cal A}\chi\ |\ {\cal A}\ |\ \psi>= <{\cal A}\chi\ |\ {\cal A}\psi>
\stackrel{antiunitarity}{=}
 <\psi\ |\ \chi>$.

By a similar argument, and because ${\cal A}^\dagger$ is also antiunitary,
one shows that one can also take ${\cal A}{\cal A}^\dagger =1$.

So, both linear unitary $\cal U$ and antilinear antiunitary $\cal A$
operators satisfy
\begin{equation}
{\cal U}{\cal U}^\dagger = 1 = {\cal U}^\dagger{\cal U},\quad
{\cal A}{\cal A}^\dagger = 1 = {\cal A}^\dagger{\cal A}.
\label{eq:ua}
\end{equation}
%

\section{Classical versus quantum Lagrangian; 
complex versus hermitian conjugation}
\label{section:cchermit}

In most literature, a fermionic Lagrangian (specially for neutrinos), is
completed by its complex conjugate.
This is because, at the classical level, a Lagrangian is a scalar and
 the fields in there are classical fields, not operators.

However, when fields are quantized, they become operators, so does the
Lagrangian which is a sum of (local) products of fields, such that, in this
case, the complex conjugate should be replaced by the hermitian conjugate.

Consider for example two Dirac fermions
$\chi = \left(\begin{array}{c} \xi^\alpha \cr
\eta_{\dot\beta}\end{array}\right)$ and
$\psi = \left(\begin{array}{c} \varphi^\alpha \cr
\omega_{\dot\beta}\end{array}\right)$;
a typical mass term in a classical Lagrangian reads
$\overline{\chi_L}\psi_R = (\xi^\alpha)^\ast \omega_{\dot\alpha} =
\xi^{\dot\alpha}\omega_{\dot\alpha} = -\omega_{\dot\alpha}\xi^{\dot\alpha}
= \omega^{\dot\alpha}\xi_{\dot\alpha}$, where we have supposed that $\xi$
and $\omega$ anticommute; its complex conjugate reads then
$(\overline{\chi_L}\psi_R)^\ast = \omega^\alpha\xi_\alpha =
(\omega^{\dot\alpha})^\ast \xi_\alpha$.

If we now consider operators
$(\overline{\chi_L}\psi_R) = [\xi^\alpha]^\dagger [\omega_{\dot\alpha}]
=[\chi_L]^\dagger [\psi_R]$,
and its hermitian conjugate is $[\omega_{\dot\alpha}]^\dagger [\xi^\alpha]
= [\omega_{\dot\alpha}^\ast][\xi^\alpha]$. Since $\left([\chi_L]^\dagger
[\psi_R]\right)^\dagger = [\psi_R]^\dagger [\chi_L]$, it only
`coincides'' with the classical complex conjugate if we adopt the
convention
\begin{equation}
\psi_R^\dagger \chi_L = (\omega^{\dot\beta})^\ast \xi_\beta,
\label{eq:hconv}
\end{equation}
where one has raised the index of $\omega$ and lowered the one of $\xi$.
We will hereafter adopt (\ref{eq:hconv}).

\section{On the use of effective expressions for the $\boldsymbol{P}$,
$\boldsymbol{C}$ and $\boldsymbol{T}$ operators
when acting on a Dirac fermion}
\label{section:becareful}

In the body of this paper we have chosen to work with fundamental Weyl fermions $\xi^\alpha$ and $\eta_{\dot \alpha}$. In order to determine 
how the discrete symmetries $P$, $C$ and $T$ act on them,
we started by their action on Dirac fermions in terms of
$\gamma$ matrices, from which, then, we deduced how each component
transforms.\\
However, one must be very cautious concerning the way
$P$, $C$ and $T$ act in terms of Dirac $\gamma$ matrices; this 
notation can  indeed easily cause confusion and induce into error,
as we show below.
It can be specially misleading when calculating the action of various
products of these three transformations and only an extreme care
can prevent from going astray. This is  why,
in manipulating these symmetry operators, we take as a general principle
to strictly  use  their  action on Weyl fermions, together with
the knowledge of their linearity or antilinearity.\\     
Since, nevertheless,  the Dirac formalism is of very common use
among physicists, we also give in the following the correct rules 
for manipulating, in this framework, discrete transformations and their
products.

Let $K$ be a transformation acting as follows
on a Dirac fermion $\psi_D$ : $K \cdot \psi_D = U_K \psi_D^{(\ast)}$,
where $U_K$ is a matrix which is in general unitary.
In the case of the usual transformations $P$, $C$ and $T$, $U_K$ may be
expressed in terms of $\gamma$ matrices. One must however keep in
mind that this does not provide a complete characterization of
the corresponding transformation, but only an effective one that must
be handled with extreme care.  It can indeed be be misleading, specially
if one relies on  ``intuition'' to infer from this expression
the linearity or antilinearity of the transformation under consideration.
This is what we showed in subsection \ref{subsec:C-lin} concerning charge
conjugation. 
Indeed,
$P \cdot \psi_D = i \gamma^0 \psi_D$ and $P$ is {\em linear} (unitary);
$C \cdot \psi_D =  \gamma^2 \psi_D^\ast$ and $C$ is {\em linear} (unitary);
$T \cdot \psi_D = i \gamma^3 \gamma^1 \psi_D^\ast$ and $T$ is
{\em antilinear} (antiunitary);
$PCT \cdot \psi_D = - \gamma^0 \gamma^1 \gamma^2 \gamma^3 \psi_D$ and $PCT$
is {\em antilinear} (antiunitary).

To illustrate this, let us investigate three \textit{a priori} possible
ways of computing the action of $PCT$, and compare them with the correct
result, obtained by applying directly to Weyl fermions the three
transformations successively (taking into account the
linear or antilinear character of operators):\newline 
* the crudest way  consists in basically multiplying the $U_K$'s, without
considering any action on a spinor (hence neglecting any
consideration concerning complex conjugation);\newline
* the second one \cite{Landau}, that we call ``Landau'' uses as a rule
the composition of the symmetry actions on a Dirac spinor;\newline
* the third one consists of making use of the linearity/antilinearity of
each transformation to move the corresponding operator through any factor
that may be present on the left of the fermion until it acts on the fermion
itself.
This last method, as we will see by going back to the 
transformation  of each component of $\psi$,
is the only correct one. 
\begin{itemize}
\item crude : $PCT\cdot \psi_D = U_PU_CU_T\psi_D
= (i \gamma^0) \gamma^2 (i \gamma^3 \gamma^1) \psi_D
 = - \gamma^0 \gamma^1 \gamma^2 \gamma^3 \psi_D.$
\item ``Landau'' : $PCT\cdot\psi_D = P\cdot(C\cdot(T\cdot\psi_D))
 = i \gamma^0 (\gamma^2(i \gamma^3 \gamma^1 \psi^\ast)^\ast)\psi_D
 = \gamma^0 \gamma^1 \gamma^2 \gamma^3 \psi_D$.
\item cautious : 
\begin{eqnarray} \nonumber
\psi_D & \stackrel{T}{\longrightarrow} & T \cdot \psi_D = i \gamma^3
\gamma^1 \psi_D^\ast \cr
& \stackrel{C}{\longrightarrow} &  C \cdot (i \gamma^3 \gamma^1 \psi_D^\ast)
\stackrel{C\ linear}{=} i \gamma^3 \gamma^1 C \cdot \psi_D^\ast
\stackrel{(\ref{eq:conjwave})}{=} i \gamma^3 \gamma^1 (C \cdot \psi_D)^\ast 
= i \gamma^3 \gamma^1 (\gamma^2)^\ast \psi_D\cr
&& \hskip 10cm = - i \gamma^3 \gamma^1 \gamma^2 \psi_D \cr
& \stackrel{P}{\longrightarrow} & P \cdot (- i \gamma^3 \gamma^1 \gamma^2
\psi_D) \stackrel{P\ linear}{=} - i \gamma^3 \gamma^1 \gamma^2 P \cdot
\psi_D 
= - i \gamma^3 \gamma^1 \gamma^2 (i \gamma^0 \psi_D)
= \gamma^3 \gamma^1 \gamma^2 \gamma^0 \psi_D\cr
 && \hskip 10cm= - \gamma^0 \gamma^1 \gamma^2 \gamma^3 \psi_D. 
\end{eqnarray}
\end{itemize} 
Similarly, when calculating the action of $(PCT)^2$, one gets:
\begin{itemize}
\item crude : $(PCT)^2\psi_D = (- \gamma^0 \gamma^1 \gamma^2 \gamma^3)
(- \gamma^0 \gamma^1 \gamma^2 \gamma^3)\psi_D = -\psi_D.$
\item ``Landau'' : $(PCT)^2\cdot \psi_D = PCT\cdot(PCT\cdot\psi_D) = (\gamma^0 \gamma^1
\gamma^2 \gamma^3)(\gamma^0 \gamma^1 \gamma^2 \gamma^3) \psi_D = - \psi_D$.
\item cautious : 
\begin{eqnarray} \nonumber
(PCT)^2 \cdot \psi_D & = & (PCT) \cdot ((PCT) \cdot \psi) \cr
& = & (PCT)\cdot(- \gamma^0 \gamma^1 \gamma^2 \gamma^3 \psi_D) \cr
& \stackrel{PCT\ antilinear}{=} & (- \gamma^0 \gamma^1 \gamma^2 \gamma^3)^\ast (PCT) \cdot \psi_D \cr
& = & (-\gamma^0 \gamma^1 \gamma^2 \gamma^3)^\ast (- \gamma^0 \gamma^1
\gamma^2 \gamma^3) \psi_D \cr
& = & \psi_D.
\end{eqnarray}
\end{itemize}
The ``cautious'' method is the only one which agrees with that
directly inferred from transforming directly Weyl spinors according to the 
rules given in the core of the paper.
One nevertheless gets the correct sign for $PCT$ (though not for $(PCT)^2$)
by the crude calculation. So, in order to discriminate without
 any ambiguity between the three ways of manipulating the symmetry operators
when acting on a Dirac fermion, {\em i.e.} to avoid (or minimize) any
risk of accidental agreement due to the cancellation of two mistakes,
we  calculated the other possible products of two operators,
and compared the results with the (reliable) ones obtained when acting
 directly on  Weyl fermions.  The results are summarized below :

\bigskip

\begin{tabular}{|c|c|c|c|}
\hline & $TP$ & $TC$ & $CP$ \\
\hline Crude (trivial product of $U$'s) & $\xi^\alpha \to -(\eta^{\dot \alpha})^\ast$ & $\xi^\alpha \to -\eta_{\dot \alpha}$ & $\xi^\alpha \to -(\xi_\alpha)^\ast$ \\
& $\eta_{\dot \alpha} \to (\xi_\alpha)^\ast$ &  $\eta_{\dot \alpha} \to \xi^\alpha$ &  $\eta_{\dot \alpha} \to -(\eta^{\dot \alpha})^\ast$ \\
& $PT = TP$ & $CT = TC$ & $PC = CP$ \\ 
\hline ``Landau'' (composition) & $\xi^\alpha \to (\eta^{\dot \alpha})^\ast$ & $\xi^\alpha \to \eta_{\dot \alpha}$ & $\xi^\alpha \to (\xi_\alpha)^\ast$ \\ 
&  $\eta_{\dot \alpha} \to -(\xi_\alpha)^\ast$ &  $\eta_{\dot \alpha} \to - \xi^\alpha$ & $\eta_{\dot \alpha} \to (\eta^{\dot \alpha})^\ast$ \\
& $PT = - TP$ & $CT = TC$ & $PC = CP$ \\
\hline Cautious (our way of computing) & $\xi^\alpha \to (\eta^{\dot \alpha})^\ast$ & $\xi^\alpha \to -\eta^{\dot \alpha}$ & $\xi^\alpha \to (\xi_\alpha)\ast$ \\
& $\eta_{\dot \alpha} \to -(\xi_\alpha)^\ast$ & $\eta_{\dot \alpha} \to \xi^\alpha$ & $\eta_{\dot \alpha} \to (\eta^{\dot \alpha})^\ast$ \\
& $PT = TP$ & $CT = - TC$ & $PC = CP$ \\
\hline 
\hline Correct result (acting directly on Weyl fermions) & $\xi^\alpha \to (\eta^{\dot \alpha})^\ast$ & $\xi^\alpha \to -\eta^{\dot \alpha}$ & $\xi^\alpha \to (\xi_\alpha)\ast$ \\
& $\eta_{\dot \alpha} \to -(\xi_\alpha)^\ast$ & $\eta_{\dot \alpha} \to \xi^\alpha$ & $\eta_{\dot \alpha} \to (\eta^{\dot \alpha})^\ast$ \\
& $PT = TP$ & $CT = - TC$ & $PC = CP$ \\
\hline
\end{tabular}
\bigskip

Moreover, our way of computing ensures that $T^2 = 1$,
in agreement with the result obtained when acting directly on Weyl spinors, 
while one faces problems with the Landau method which leads to
$T^2 = -1$.
Indeed, 
$T^2 \cdot \psi_D = T \cdot (i \gamma^3 \gamma^1 \psi_D^\ast)
\stackrel{T\ antilinear}{=} -i \gamma^3 \gamma^1 T \cdot \psi_D^\ast 
\stackrel{(\ref{eq:conjwave})}{=} -i \gamma^3 \gamma^1 (T \cdot \psi_D)^\ast
= -i \gamma^3 \gamma^1 (-i)
\gamma^3 \gamma^1 \psi_D = \psi_D$, 
while ``Landau's'' prescription leads to
$T^2 \cdot \psi_D = i \gamma^3 \gamma^1 (i \gamma^3 \gamma^1
\psi_D^\ast)^\ast = i \gamma^3 \gamma^1 (-i) \gamma^3 \gamma^1 \psi_D 
= \gamma^3 \gamma^1 \gamma^3 \gamma^1 \psi_D = - \psi_D.$

\newpage\null
\begin{em}

\end{em}

\end{document}